\documentclass[numberedappendix]{emulateapj}
\usepackage{apjfonts, natbib}
\usepackage{rotating}

\newcommand{\be}{\begin{equation}}
\newcommand{\ee}{\end{equation}}

\newcommand{\I}{\it{I}\rm}
\newcommand{\Q}{\it{Q}\rm}
\newcommand{\U}{\it{U}\rm}
\newcommand{\Qf}{$Q/I$}
\newcommand{\Uf}{$U/I$}
\newcommand{\flow}{100~GHz}
\newcommand{\fhigh}{150~GHz}

\shorttitle{QUaD Galactic Plane Survey}
\shortauthors{QUaD collaboration}

\begin{document}

\slugcomment{Submitted to ApJ}

\title{The QUaD Galactic Plane Survey 1: Maps And Analysis of Diffuse Emission}

\author{
  QUaD collaboration
  --
  T.\,Culverhouse\altaffilmark{1},
  P.\,Ade\altaffilmark{2},
  J.\,Bock\altaffilmark{3,4},
  M.\,Bowden\altaffilmark{2,5},
  M.\,L.\,Brown\altaffilmark{7},
  G.\,Cahill\altaffilmark{8},
  P.\,G.\,Castro\altaffilmark{6,9},
  S.\,Church\altaffilmark{5},
  R.\,Friedman\altaffilmark{1},
  K.\,Ganga\altaffilmark{10},
  W.\,K.\,Gear\altaffilmark{2},
  S.\,Gupta\altaffilmark{2},
  J.\,Hinderks\altaffilmark{5,11},
  J.\,Kovac\altaffilmark{12},
  A.\,E.\,Lange\altaffilmark{4},
  E.\,Leitch\altaffilmark{3,4},
  S.\,J.\,Melhuish\altaffilmark{2,13},
  Y.\,Memari\altaffilmark{6},
  J.\,A.\,Murphy\altaffilmark{8},
  A.\,Orlando\altaffilmark{2,4}
  R.\,Schwarz\altaffilmark{1},
  C.\,O'\,Sullivan\altaffilmark{8},
  L.\,Piccirillo\altaffilmark{2,13},
  C.\,Pryke\altaffilmark{1},
  N.\,Rajguru\altaffilmark{2,14},
  B.\,Rusholme\altaffilmark{5,15},
  A.\,N.\,Taylor\altaffilmark{6},
  K.\,L.\,Thompson\altaffilmark{5},
  A.\,H.\,Turner\altaffilmark{2},
  E.\,Y.\,S.\,Wu\altaffilmark{5}
  and
  M.\,Zemcov\altaffilmark{2,3,4}
}

\altaffiltext{1}{Kavli Institute for Cosmological Physics,
  Department of Astronomy \& Astrophysics, Enrico Fermi Institute, University of Chicago,
  5640 South Ellis Avenue, Chicago, IL 60637, USA.}
\altaffiltext{2}{School of Physics and Astronomy, Cardiff University,
  Queen's Buildings, The Parade, Cardiff CF24 3AA, UK.}
\altaffiltext{3}{Jet Propulsion Laboratory, 4800 Oak Grove Dr.,
  Pasadena, CA 91109, USA.}
\altaffiltext{4}{California Institute of Technology, Pasadena, CA
  91125, USA.}
\altaffiltext{5}{Kavli Institute for Particle Astrophysics and
Cosmology and Department of Physics, Stanford University,
382 Via Pueblo Mall, Stanford, CA 94305, USA.}
\altaffiltext{6}{Institute for Astronomy, University of Edinburgh,
  Royal Observatory, Blackford Hill, Edinburgh EH9 3HJ, UK.}
\altaffiltext{7}{Cavendish Laboratory,
  University of Cambridge, J.J. Thomson Avenue, Cambridge CB3 OHE, UK.}
\altaffiltext{8}{Department of Experimental Physics,
  National University of Ireland Maynooth, Maynooth, Co. Kildare,
  Ireland.}
\altaffiltext{9}{{\em Current address}: CENTRA, Departamento de F\'{\i}sica,
  Edif\'{\i}cio Ci\^{e}ncia, Piso 4,
  Instituto Superior T\'ecnico - IST, Universidade T\'ecnica de Lisboa,
  Av. Rovisco Pais 1, 1049-001 Lisboa, Portugal.}
\altaffiltext{10}{APC/Universit\'e Paris 7  Denis Diderot/CNRS, B\^atiment Condorcet,
  10, rue Alice Domon et L\'eonie Duquet, 75205 Paris Cedex 13, France.}
\altaffiltext{11}{{\em Current address}: NASA Goddard Space Flight
  Center, 8800 Greenbelt Road, Greenbelt, Maryland 20771, USA.}
\altaffiltext{12}{Harvard Department of Astronomy, 60 Garden St, Cambridge MA 02138, USA.}
\altaffiltext{13}{{\em Current address}: School of Physics and
  Astronomy, University of
  Manchester, Manchester M13 9PL, UK.}
\altaffiltext{14}{{\em Current address}: Department of Physics and Astronomy, University
  College London, Gower Street, London WC1E 6BT, UK.}
\altaffiltext{15}{{\em Current address}:
  Infrared Processing and Analysis Center,
  California Institute of Technology, Pasadena, CA 91125, USA.}

\begin{abstract}

We present a survey of $\sim800$ square degrees of the galactic plane observed 
with the QUaD telescope. 
The primary product of the survey are maps of Stokes \I, \Q~and \U~parameters
 at 100 and \fhigh, with spatial resolution 5 and 3.5 arcminutes respectively.
Two regions are covered, spanning approximately $245-295^\circ$ and $315-5^\circ$
 in galactic longitude $l$, and $-4<b<+4^\circ$ in galactic latitude $b$.
At $0.02^{\circ}$ square pixel size, the median sensitivity is 74 and 107 kJy/sr
 at \flow~and \fhigh~respectively in \I, and 98 and 120 kJy/sr for \Q~and \U.
In total intensity, we find an average spectral index of 
$\alpha=2.35\pm0.01\mathrm{(stat)}\pm0.02\mathrm{(sys)}$ for $|b|\leq1^\circ$, 
indicative of emission components other than thermal dust.
A comparison to published dust, synchrotron and free-free models implies an 
excess of emission in the \flow\ QUaD band, while better agreement is found at \fhigh.
A smaller excess is observed when comparing QUaD \flow\ data to WMAP 5-year
W band; in this case the excess is likely due to the wider bandwidth of QUaD.
Combining the QUaD and WMAP data, a two-component spectral fit to 
the inner galactic plane ($|b|\leq1^{\circ}$) yields mean spectral indices of
$\alpha_{s}=-0.32\pm0.03$ and $\alpha_{d}=2.84\pm0.03$; the former is 
interpreted as a combination of the spectral indices of synchrotron, free-free
and dust, while the second is attributed largely to the thermal dust
continuum.
In the same galactic latitude range, the polarization data show a high degree
 of alignment perpendicular to the expected galactic magnetic field direction, 
and exhibit mean polarization fraction $1.38\pm0.08\mathrm{(stat)}\pm0.1\mathrm{(sys)}$\% 
at \flow\ and $1.70\pm0.06\mathrm{(stat)}\pm0.1\mathrm{(sys)}$\% at \fhigh.
We find agreement in polarization fraction between QUaD \flow~ and WMAP W band, 
the latter giving $1.1\pm0.4$\%.

\end{abstract}
\keywords{Surveys --- Submillimeter ---
          cosmic microwave background polarization ---
          cosmology: observations --- diffuse radiation --- 
          Galaxy: structure --- Galaxy: disk --- ISM: structure}

\section{Introduction}

%
%
%
%
%
%
%
%
%
%
%
%
%

Radio and sub-mm observations of the galactic plane yield important insights
into many astrophysical processes associated with galaxies like our own, from the large 
scale properties of magnetic fields, to smaller scale phenomena associated 
with star formation.

These properties can often be inferred from low resolution observations of diffuse galactic
components, or large samples of representative objects distributed through the
galaxy.
Three mechanisms contribute to the diffuse galactic emission in total intensity in 
the radio and sub-mm:
Synchrotron radiation, which dominates below $\sim60~\mathrm{GHz}$ and is caused
by relativistic electrons spiralling in magnetic fields; free-free
emission, which is generated by non-relativistic electron-ion interactions; and radiation
from vibrational modes of thermal dust, whose emission dominates above $\sim100~\mathrm{GHz}$.
Of these, synchrotron and dust are appreciably polarized, with typical polarization
fractions close to the galactic plane of $\sim2-4\%$ and $\sim1\%$ respectively~\citep{kogut07}; both
result in polarized light aligned perpendicular to the magnetic field.

The polarization of dust is due to prolate grains aligning with their long axis 
perpendicular to the local magnetic field~\citep{lazarian2003}, 
with the polarization fraction dependent on the grain size distribution and their 
overall alignment~\citep[e.g.][]{prunet98}.
Observations of polarized starlight via dust absorption have indirectly demonstrated
 a large degree of coherence of the magnetic field in our galaxy and 
others~\citep{heiles1996,zweibel1997}. 
However, these measurements can be biased by lines of sight with low column densities~\citep{benoit04}
and therefore more direct probes of the galactic dust are desirable, not only for the
study of dust in its own right, but as a probe of the magnetic field responsible 
for most large-scale polarized emission in the galaxy~\citep[e.g.][]{hildebrand2000}.
Sub-mm polarization vectors can be reasonable tracers of the magnetic field
structure even for relatively dense clouds, and are therefore an attractive option
for this line of study.

The emissive properties of dust in the sub-mm have also attracted 
attention from the CMB community, since diffuse galactic polarization poses a 
challenging obstacle to the detection of primordial gravitational 
waves via the `B-mode' polarization signal~\citep[e.g.][]{hu_white97,Dunkley2008}.
As the B-mode power spectrum is predicted to peak on angular scales 
$\sim1^\circ$, characterization of dust as a foreground is essential 
to account for this `contaminant' from surveys over large areas of sky, 
such as that expected from the Planck satellite~\citep{Planck}.

Observational constraints on diffuse galactic polarization from dust are currently 
limited to a small number of experiments, including WMAP from $23$--$94$~GHz at 
resolution up to $\sim0.25^{\circ}$~\citep{kogut07,gold08}, and 
Archeops at $353$~GHz smoothed to $\sim1^\circ$ 
resolution~\citep{benoit04,ponthieu05}.
While a variety of models exist for estimating the unpolarized 
contribution of dust~\citep[e.g.][hereafter FDS]{finkbeiner99}, the limited 
number of experiments at dust-dominated frequencies has prevented 
detailed comparison to observations.
Furthermore, the lack of angular resolution of such experiments means emission
from diffuse and discrete sources cannot be separated, particularly in the plane 
of the galaxy ($|b|<10^\circ$).

The study of discrete sources at and above $100$~GHz yields insights into the process of star formation.
In star forming regions, thermal dust efficiently absorbs UV 
light from star formation, and re-radiates in the sub-mm where dust is optically thin.
Observations near the spectral peak can therefore probe the centers 
of dense cores and constrain the stellar core mass 
function~\cite[e.g.][]{netterfield2009,schuller2009,olmi2009}.

In this paper we report an $\sim800$ square degree survey of 
the galactic plane with the QUaD telescope, which operated at 
100 and \fhigh~with angular resolution of $5'$ and $3.5'$ respectively,
 in Stokes \I, \Q~and \U~parameters
.
A survey of this size, frequency and angular resolution can be used to 
investigate the polarized and unpolarized properties of both diffuse emission
and discrete sources.
The mapmaking and properties of the diffuse emission form the core of this paper; a 
companion publication (hereafter the ``Source Paper'' --- Culverhouse et al. in 
prep.) contains analysis of the compact source distribution in the survey. 

\section{Instrument Summary and Observations}
\label{sec:instobs}

Here we summarize the features of the QUaD experiment~--- a detailed
description can be found in~\citep{hinderks08}, hereafter referred 
to as the ``Instrument Paper''. 
QUaD was a 2.6~m Cassegrain radio telescope on the mount originally 
constructed for the DASI experiment~\citep{leitch02}, and enclosed 
in an extended reflective ground shield.
The receiver consisted of 31~pairs of polarization sensitive
bolometers or PSBs~\citep{2003SPIE.4855..227J}, 12 at \flow, and 19 at
 \fhigh, with each PSB pair located within a single feed.
The PSB pairs were split into two orientation groups to allow simultaneous
measurement of Stokes \Q~and \U.
QUaD operated from February 2005 to November 2007; the observations 
reported in this paper were taken over 40 days between July and October
 2007. 

\begin{figure}[t]
\resizebox{\columnwidth}{!}{\includegraphics{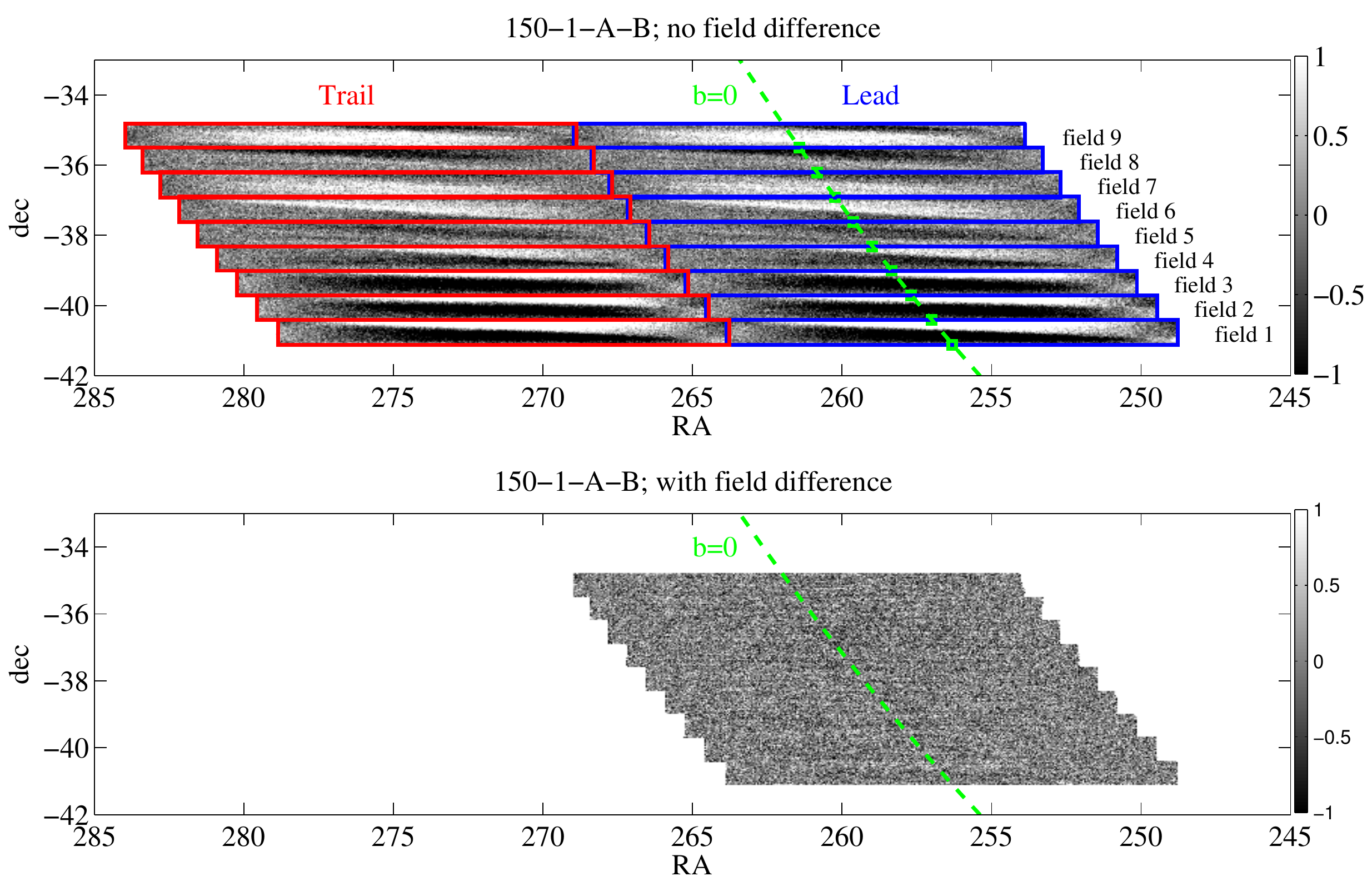}}
\caption{{\it Top:} Coadded PSB pair difference map of a single \fhigh~
feed from one day of observations, with no field difference. 
The nine rectangular lead-trail pairs of fields observed on this 
day are shown in blue and red respectively.
Note the strongly polarized ground pickup which repeats between lead-trail 
pairs. 
The colour scale is MJy/sr, with the green line indicating the plane of 
the galaxy.
Green squares indicate the starting point of the scan pattern for each field.
{\it Bottom:} As above, but with field differencing applied --- the ground
signal is now heavily suppressed.}
\label{fig:fielddiff1}
\end{figure}

As with the QUaD CMB observations~\citep{ade07,QUAD08,QUAD09}, the second 
of which we refer to as ``P09'', a lead-trail field differencing 
scheme was employed to allow subtraction of contaminating ground pickup.
Each day, nine lead-trail pairs of fields were observed, with the lead 
field tracking center aligned with the plane of the galaxy $b=0$.
For each lead field, the companion trail field repeated exactly the 
same azimuth (az) and elevation (el) scan pattern with respect to the ground,
but with the tracking center 1hr later in RA. 
In order to minimize the possibility of temporal variation in ground
 signal, each trail field was observed immediately after its 
companion lead field, resulting in a lead 1, trail 1, lead 2, trail 2 
etc ordering --- see Figure~\ref{fig:fielddiff1} for a plot of this scheme.
Though the signal levels in trail fields are generally much smaller
than the lead field, some spurious signal is introduced into the lead field
due to field-differencing.
A discussion of this effect is presented in Appendix~\ref{subsec:processeffects}, 
while Appendices~\ref{subsec:specindrecovery} and~\ref{subsec:polfracrecovery} 
demonstrate that the recovery of global parameters of the diffuse 
emission are unaffected by field-differencing or filtering.

Whilst tracking each field center, the telescope scanned forward and 
backward in az with a total az range of 15$^{\circ}$ at a rate of 0.4$^\circ$/s, 
followed by a step of -0.02$^\circ$ in elevation. 
This was repeated 35 times to build up a raster map of the sky, before
 slewing to the next field where the same scanning pattern was repeated. 
Each field took approximately 1hr observing time, and covered 
0.7$^\circ$ in dec, a total of 6.3$^\circ$ per day.
Figure~\ref{fig:fielddiff1} shows a graphic representation of the
scanning strategy, and demonstrates that ground signal is cleanly 
removed by field-differencing.

\begin{figure*}[ht]
\resizebox{\textwidth}{!}{\includegraphics{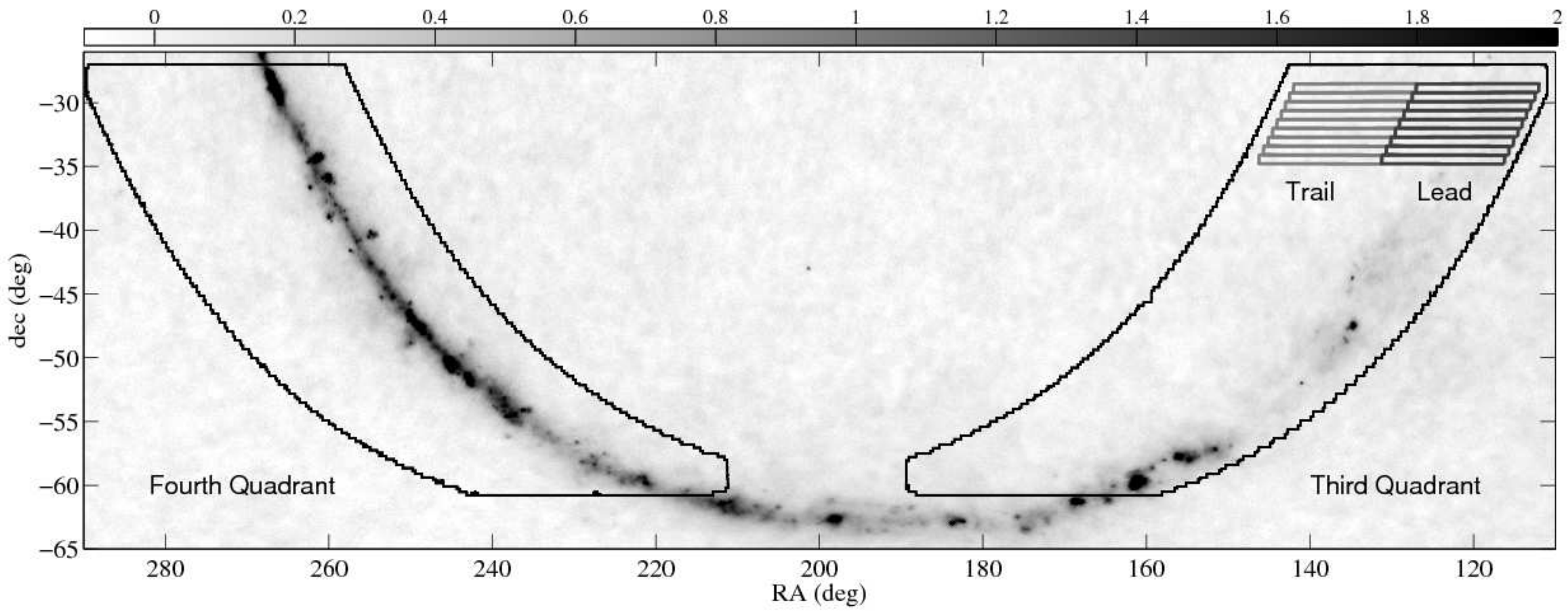}}
\caption{Total QUaD galactic survey area at \fhigh\ (black lines) plotted 
over WMAP 5yr W band total intensity map~\citep{hinshaw2009}.
The color scale is in MJy/sr.
For one day of data, lead field coordinates for center pixel are plotted 
in dark gray; trail fields are in light gray. 
This lead-trail observing strategy is repeated over the rest of the survey
region within the black locus.}
\label{fig:surveyarea}
\end{figure*}

The entire survey coverage is shown in Figure~\ref{fig:surveyarea}, and was 
limited in declination range by two factors. 
First, the beams intersect the ground shield at 
elevations lower than $\sim25^\circ$ ($\mathrm{dec}>-25^\circ$).
Second, at $\mathrm{dec}<-60^\circ$ the galactic plane is nearly parallel to the
horizon --- filtering the scans, necessary to remove atmospheric contamination, 
would remove the bulk of the diffuse galactic emission unless the scan 
length was massively increased, resulting in a loss of sensitivity.

Two portions of the galaxy are available between $-60$ and~$-25^\circ$ in 
dec, between $110^\circ<\mathrm{RA}<190^\circ$ and $210^\circ<\mathrm{RA}<290^\circ$. 
In galactic longitude and latitude ($l$,$b$) these correspond to a 
latitude range approximately $-4^\circ< b< 4^\circ$, and 
$240^\circ< l< 295^\circ$ and $315^\circ< l< 5^\circ$ --- hereafter, these
two regions are loosely termed the `third quadrant' and `fourth quadrant'.
With our coordinate constraints, the third and fouth quadrant regions each 
spanned~$32^\circ$ in dec; the 40 days of observations were equally 
divided between the two regions, allowing four complete passes over each.

\section{Low Level Data Processing}
\label{sec:lldataproc}

 Low-level processing of the raw timestream is performed using the same steps
 described in P09.
The timestream is deconvolved to remove the effect of the bolometer 
timeconstants, deglitched, and downsampled.
Between observations of each field, relative gains between PSB pairs in a 
feed, and between feeds in each frequency group, are determined using `el-nods'. 
In this calibration procedure, the telescope is moved first up then down by 
one degree in elevation --- this injects a large signal into the timestream 
due to the common-mode atmospheric gradient; further details are given in the
 Instrument Paper.

As seen in Figure~\ref{fig:fielddiff1}, the ground pickup is strong compared
to the sky signal of interest.
A misalignment between absolute lead and trail field azimuth coordinates 
$\delta\mathrm{az}$ causes ground signal to cancel imperfectly, and can lead to 
significant contamination for $\delta\mathrm{az}>0.1^\circ$.
This is carefully corrected by using the pointing information to realign the
lead and trail fields on a scan-by-scan basis, at the expense of a small
quantity of data where there is no overlap between lead and trail scans.
Approximately one out of nine lead-trail pairs require correction of
up to $0.4^\circ$, leading to a loss of data of order 0.1\%.

Further data is rejected from visual inspection of field-differenced maps, 
made using sum and difference data from each feed.
Maps of the same fields taken on different days are compared 
to distinguish the repeatable sky signal from spurious contamination.
The pair difference maps in particular are useful because the amplitude of
the contamination, which appears as spurious $1/f$ noise, is considerably 
larger than that of the polarized galactic signal.
Typical rejection rates are one out of nine fields for four bolometer pairs,
a loss of data of $\sim0.75\%$.

\section{Mapmaking}
\label{sec:mapmaking}

The map-making scheme used here is a multi-stage adaptation of the `naive'
mapping used in P09, and requires information on telescope pointing (both 
absolute and the relative offsets of each PSB in the focal plane), and the
 PSB angles and efficiencies to construct the Stokes \I, \Q~and U maps.
As in P09 and unless stated otherwise, the pixelization for all
maps in this paper is in RA and dec, using 0.02$^\circ$ square pixels.
Further details of constructing \I, \Q~and U maps from timestream may
be found in P09; here we summarize the basic points.

\subsection{Pointing, PSB Angles and Efficiencies}
\label{sec:pointangles}

Absolute pointing was determined from a nine parameter online pointing model,
derived from optical and radio observations as described in the Instrument Paper.
This was shown to have an absolute accuracy over the hemisphere of $\sim0.5'$ 
rms from pointing checks on RCW38 and other galactic sources, taken over 
two seasons of CMB observations.

The scatter in the centroid positions for a given detector relative to the
boresight was consistent with the overall pointing wander of $\sim0.5'$ rms, 
and the offset angles of each detector around the focal plane showed no 
evidence for systematic changes with time.
The detector offset angles used in mapmaking are the mean of the values
 observed from RCW38, with an estimated uncertainty of $\sim0.15'$.

PSB polarization angles and efficiencies were determined using a 
chopped thermal source placed behind a polarizing grid and observed
at many angles --- further details are given in P09 and the Instrument
Paper. From these observations we also measure our mean cross polar 
leakage (the response of a single PSB to anti-aligned radiation) as 
$\epsilon=0.08\pm0.015$. This mean value is applied to all channels
when constructing maps, and all simulations include the scatter about
the mean. 

For an experiment of this type cross polar leakage
does not imply leakage from total to polarized intensity --- it is
 simply a small loss of efficiency, which is corrected by an 
additional calibration factor applied to the polarization data.

\subsection{Mapmaking Algorithm}
\label{sec:mapmethod}

Before binning into maps, the data is first field-differenced to
 remove ground contamination --- this operation is performed 
directly in the timestream. 
For each feed the sum and difference of the data is then taken for each 
pair of PSBs.
To construct the \I\ maps, we coadd the pair-sum data for each feed 
on each day.
For polarization, a $2\times2$ matrix inversion is required for each
 map pixel to convert from the pair difference data to \Q~and \U~maps. 
This matrix expresses the polarized sky intensity projected
 onto each PSB, which is measured in the pair-difference timestream. 
To invert such a matrix we require each pixel be measured at two PSB 
angles --- this is achieved with the two orientation groups in the QUaD 
focal plane.
However, before coadding the data into maps, timestream filtering is
 required to reduce the low-frequency noise, which causes striping in
the scan direction in the maps.

\subsubsection{Initial Filtering}
\label{subsubsec:initfilt}

In addition to detector noise, bolometer drifts and 
atmospheric $1/f$ noise are a large contribution to the sum data (since 
the atmosphere is largely unpolarized, its contribution is
common-mode and is thus heavily suppressed in the difference data).
The CMB analysis of P09 subtract a third-order polynomial from
the timestream to limit the effects of this $1/f$ noise; though this 
filtering removes sky signal, the effect is accounted for in 
simulations, which is a feasible method for a power spectrum analysis.
However, we are interested in the spatial distribution of the galactic
signal, and the timestream may not simply be filtered in the same way.
For example, in the sum data the bright galactic emission will 
dominate the polynomial fit, leading to regions of unphysical negative 
signal when the polynomial is subtracted from the data; a minimal level 
of filtering is therefore desirable. 
Here, the end portions of each scan --- the most distant parts of the scan from
the galactic plane --- are used to determine a DC level and slope, which 
is then subtracted over the entire scan.
The same procedure is used for total and polarized intensity data.
This choice of filtering scheme effectively forces the maps to be
zero at the edges, a consequence of our inability to measure the DC level
of the sky brightness.

The amount of scan ends used is a trade off between better determination
of the $1/f$ noise, which requires an increasing fraction of the scan, and 
larger regions of negative intensity in the final maps.
The second of these effects arises because the DC level subtracted from
the timestream is influenced more by the bright galactic signal as more of 
the scan is used.

After some experimentation, using a scan fraction $f_{s}=25\%$ (i.e. 
$\sim12.5\%$ at either end) was found to result in total intensity maps 
with few negative regions, while keeping residual $1/f$ noise to a minimum.
The quantitative results on the diffuse emission, presented in Section~\ref{sec:diffprop}, 
are not significantly changed by adopting a different fraction of the 
scan, or using a mask of fixed width either side of the galactic plane.

From the sum/difference data we construct a minimally-filtered map, 
termed the `initial map' or $m_{0}$: scans are filtered by removing a
 DC-level and linear slope, as described above.
The timestream is then coadded into the map using the pointing 
information for each pair of PSBs, and weighted by the inverse scan 
variance as determined from the scan end data after filtering. 
A section of the $m_{0}$ survey map is shown in the top panel of 
Figure~\ref{fig:quadfiltstages}.
Scan variances are coadded into a `variance map', which produces an estimate
of the pixel variance over the survey area.

\subsubsection{Secondary Filtering}
\label{subsubsec:secondfilt}

\begin{figure}[ht]
\resizebox{\columnwidth}{!}{\includegraphics{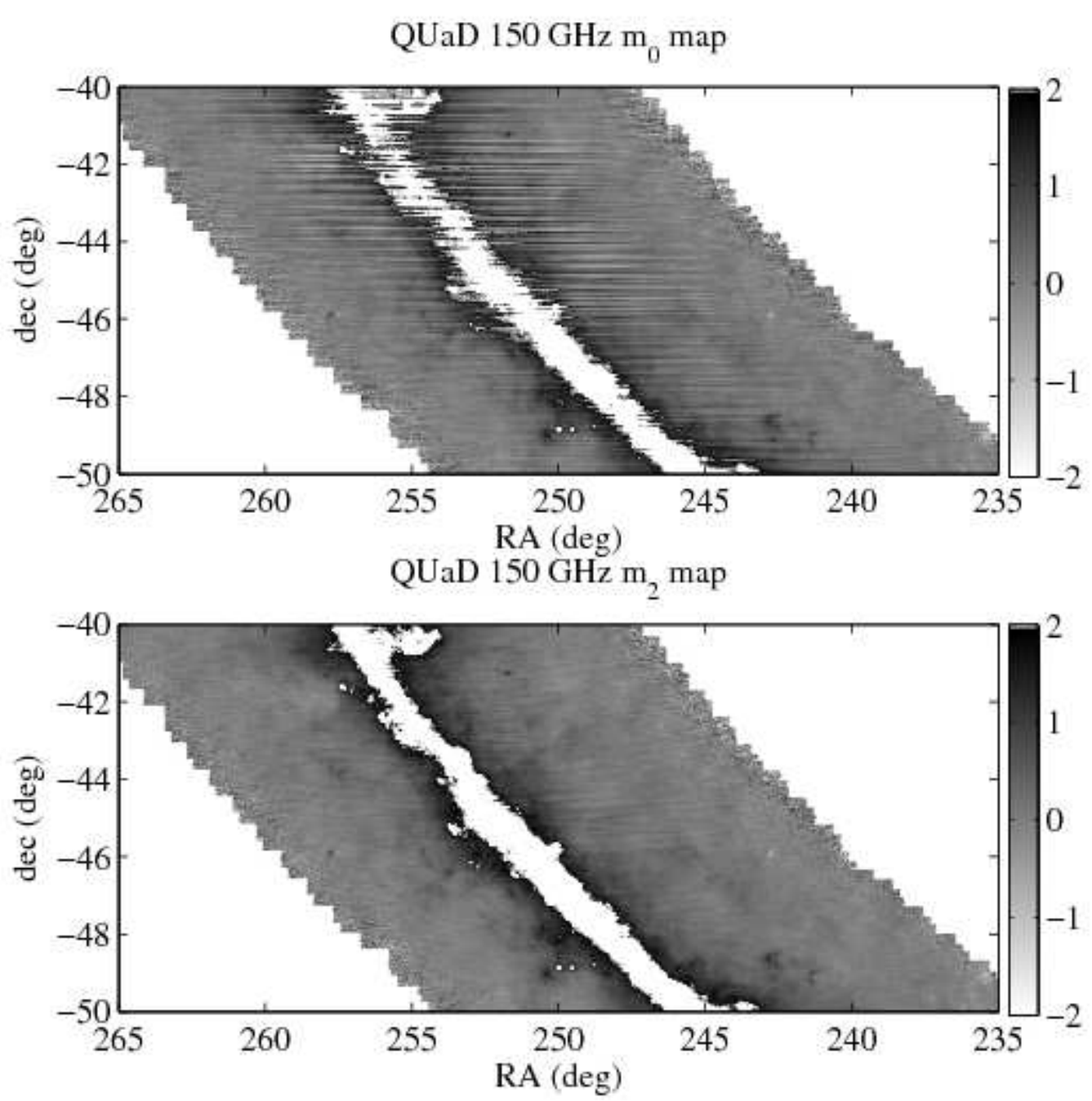}}
\caption{Maps illustrating the destriping process on QUaD \fhigh~
data.
All color scales are in MJy/sr, with saturated pixels within the survey
region shown in white.
{\it Top:} DC + slope filtered ($m_{0}$) map.
{\it Bottom:} Destriped map $m_{2}$.}
\label{fig:quadfiltstages}
\end{figure}

Small-scale noise between rows of map pixels (visible in the top panel of
Figure~\ref{fig:quadfiltstages}) is further reduced by a destriping algorithm as follows.
After the $m_{0}$ map is constructed, compact sources are located using
the source extraction code described in the Source Paper, which is based
on the SExtractor routine~\citep{bertin96}.
A second map ($m_{1}$) is generated identically to $m_{0}$, with the exception
that sources located in the scan ends are masked during filtering.
This process prevents discrete sources lying far from the bulk of the
diffuse emission from influencing the inital polynomial filtering.

The simple removal of a DC-level and slope from each scan produces a 
map which still exhibits striping due to atmospheric $1/f$ noise.
To suppress this noise, we construct a template for the sky signal,
 which is simply a smoothed version of the $m_{1}$ map.
This template is then subtracted from the raw timestream, leaving data
which is dominated by atmospheric noise.
A 6th-order polynomial is then fit to the signal-subtracted
timestream, and then subtracted from the original data.
The timestream still contains the galactic signal of interest, but with
the $1/f$ noise much suppressed compared to the simple DC+slope filtering
described above.

Maps made with the secondary-filtered data are termed the $m_{2}$ maps --- 
these exhibit improved noise properties without a heavy penalty in signal
 loss, as may be seen in Figure~\ref{fig:quadfiltstages}.
A detailed discussion of the algorithm and its implementation is 
presented in Appendix~\ref{app:destriping}.

\subsection{Absolute Calibration}
\label{subsec:abscal}

Absolute calibration is applied using conversion factors from~\cite{QUAD09}.
The maps used in the QUaD CMB analysis were cross-calibrated with
the Boomerang experiment~\citep{masi06} to produce factors
 which convert from detector units of volts to thermodynamic 
units of $\mu$K. 
These are estimated to be uncertain at the 3.5\% level.
The galactic maps are calibrated using the same conversion factors,
and then rescaled to brightness units using
\begin{equation}
dI = (dB/dT)_{2.73}\times dT,
\label{eq:dbdt}
\end{equation}
where $B$ is the Planck function, and $dT$ and $dI$ represent thermodynamic
and brightness fluctuations respectively.
Througout the analysis in this paper, the central 
frequencies $\nu_{0}$ of the QUaD bands are loosely referred to 
as 100 and \fhigh.
Assuming a spectrally flat source, numerical integration of the 
QUaD bandpass presented in the Instrument Paper results in central 
frequencies of 94.5 and 149.6 GHz.

\subsection{Final Maps}
\label{subsec:finalmaps}

\begin{figure*}[hp]
\centering
{\includegraphics[width=22cm,angle=90]{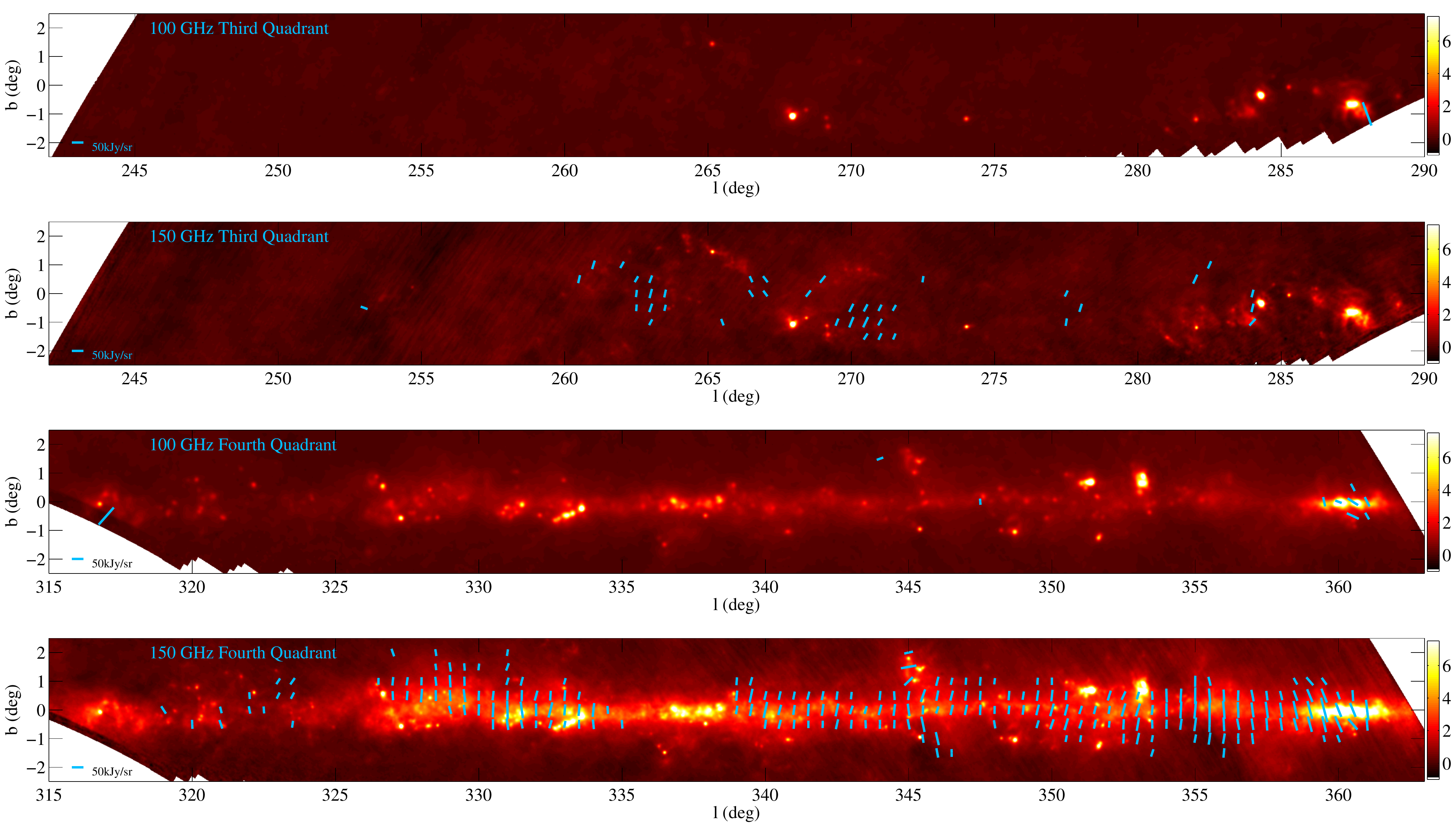}}
\caption{Total intensity and polarization vector maps
for third quadrant 100 GHz (first) and \fhigh~(second), and fourth quadrant
 100 GHz (third) and \fhigh~(fourth).
The total intensity maps are full-resolution maps smoothed to the beam 
scale; the color scale is in MJy/sr. 
The polarization vectors are derived from coarse resolution maps with
$0.5^\circ$ pixels.
Only polarization vectors for which $P/\sigma_{P}>5$ are shown.}
\label{fig:brightmaps}
\end{figure*}

Calibrated, destriped survey maps of Stokes $I$, $Q$, and $U$ are shown in 
Appendix~\ref{app:maps}, Figures~\ref{fig:maps_T}--\ref{fig:maps_U} in 
celestial coordinates (the native coordinate system for map processing) .
The \flow~maps are hereafter referred to as $I_{100}$, $Q_{100}$, 
and $U_{100}$, and likewise at \fhigh.
All polarization maps follow the IAU convention, in 
which $+Q$ is parallel to N-S and $+U$ parallel to NE-SW~\citep{IAU96}.
Converting to galactic coordinates (galactic longitude $l$ and latitude $b$) 
results in the maps shown in Figure~\ref{fig:brightmaps}.
Total intensity maps are transformed from the native celestial coordinates to galactic $l$ 
and $b$ by linear interpolation of the map pixel values between the respective
coordinate grids calculated with standard astronomical software packages.
For polarization, we compute the angles between unit (RA,dec) vectors in the
($l$,$b$) basis at each point in the map --- these angles are then used to 
project the polarized intensity onto the ($l$,$b$) basis.

\begin{figure}[h]
\resizebox{\columnwidth}{!}{\includegraphics{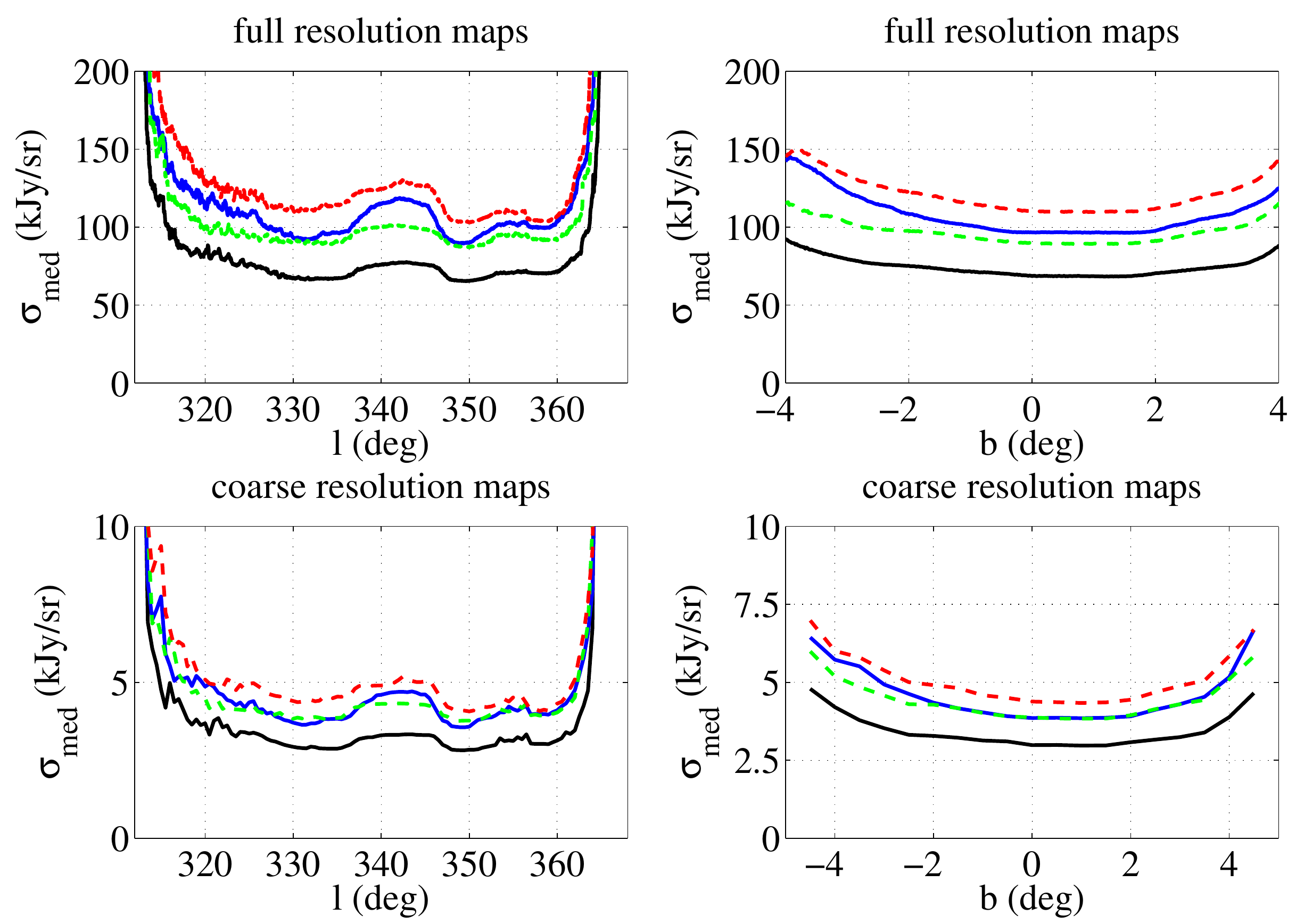}}
\caption{Plots showing median sensitivity of the survey as a function of
$l$ and $b$ for the fourth quadrant region data; plots for the third quadrant
 region are very similar. 
In all panels, black represents 100 GHz \I, blue \fhigh~\I, green 100 GHz
\Q, and red \fhigh~\Q.
The sensitivity to U is comparable to \Q.
Top row is for the full resolution (pixel size $0.02^\circ$) maps, while
bottom row is for the coarse resolution ($0.5^\circ$ pixels) maps used 
for the analysis of diffuse structure in Section~\ref{sec:diffprop}.
}
\label{fig:sensitivity}
\end{figure}


The median sensitivity of the survey as a function of $l$ and $b$ is 
plotted in Figure~\ref{fig:sensitivity} for the fourth quadrant region, where 
the rms values are determined from the variance maps; the sensitivity
is almost identical for the third quadrant.
The QUaD galactic plane survey is fairly uniform over $\sim50^\circ$ in
$l$ and $|b|\leq4^\circ$ in each region, allowing detection and 
characterization of diffuse and localized emission over $\sim800$ square
 degrees of the low-latitude galaxy.

At the native map resolution of $0.02^\circ$, much of the diffuse 
emission and hundreds of compact sources are detected to high significance; particularly
bright regions such as the galactic center reach signal-to-noise ratios 
$S/N>500$ in total intensity, with the bulk of the diffuse emission 
(within $\sim3^\circ$ of the galactic plane) detected with $S/N>10$ per map pixel.
In Figure~\ref{fig:sensitivity} we also show the sensitivity for `coarse' resolution 
maps with pixel size of $0.5^\circ$; in these maps, the equivalent 
brightness sensitivity is a factor $\sim25$ higher than the full resolution
maps. 
In polarization, there is significant diffuse polarized emission in the \fhigh~
fourth quadrant data at this resolution, while polarized signal at \flow~
becomes apparent when degrading to 0.5$^\circ$ pixels.
Note that when constructing coarse resolution maps, all timestream operations
such as field-differencing, filtering and destriping are performed exactly as 
above. 
Locating point sources, which forms part of the destriping process, is done 
on the native $0.02^\circ$ resolution $m_{0}$ and $m_{1}$ maps as above --- 
the only difference between making the coarse and full resolution maps is 
the map pixel size which the processed timestream is binned into when 
constructing the final $m_{2}$ maps.

The fainter diffuse \flow~signal implies that we are 
observing emission with a positive spectral index $\alpha$ 
($I\propto\nu^{\alpha}$), likely dominated by polarized thermal dust.
As reported in the Source Paper, a small number of discrete sources
are detected in the polarization maps, along with a polarized 
arc near the galactic center, and an extended polarized cloud.

In addition to the signal maps, we also generate `jackknife' maps
in an idential manner to P09.
The timestream data is split evenly in two ways, scan direction and time (i.e.
first half of data against second half).
For each jackknife, maps are constructed from each split of the
data exactly as for the signal maps; the difference of pairs of maps from each split
is then taken.
These jackknife maps provide useful tests of possible systematics in the
data or mapmaking process, and are generated for $I$, $Q$ and $U$ maps
at both frequencies.

\section{Properties of Diffuse Emission}
\label{sec:diffprop}

The QUaD data can be used to place constraints on interesting properties of
the bulk galactic emission, namely the total intensity spectral index, 
the polarization fraction, and the angle of polarization.
Due to the low signal-to-noise of polarization, we use the coarse resolution 
maps described above for the analysis; for convenience, the maps are 
interpolated to galactic coordinates as described in Section~\ref{subsec:finalmaps}.
On the coarse pixel grid, the effects of differing beam sizes between 
the two frequency bands become negligible compared to the effects of 
pixelization.

Figure~\ref{fig:brightmaps} shows polarization vectors from the coarse maps 
overlaid on full resolution total intensity maps for the $|b|<2.5^{\circ}$
region of the survey.
Pixels with polarized intensity signal-to-noise $P/\sigma_{P}>5$ have 
been excluded, with the polarized intensity
\begin{equation}
P^{2}=Q^{2}+U^{2}
\label{eq:totpol}
\end{equation}
and its error given by
\begin{equation}
\sigma_{P}^{2}=\sigma_{Q}^{2}+\sigma_{U}^{2}+2\sigma^{2}_{QU},
\label{eq:sigtotpol}
\end{equation}
where $\sigma_{Q}^{2}$ and $\sigma_{U}^{2}$ are the variances in \Q~and \U,
and $\sigma_{QU}$ is the covariance between \Q~and \U~map pixels.
The \fhigh~ data in particular show a high degree of alignment between 
the polarization (pseudo) vectors.
In the statistical analysis that follows, pixels from the third and fourth
quadrant maps are combined.
Throughout, we ignore the contribution of primordial CMB anisotropies; the bulk of the
analysis is performed within $-1^{\circ}<b<1^{\circ}$, where the galactic emission is
expected to be overwhelmingly dominant in both total and polarized intensity.
In addition, pixels whose polarized emission is dominated by discrete sources rather than 
the diffuse background are also excluded.
This flagging only includes the galactic center and a polarized cloud at 
$l\sim345^\circ$, $b\sim1.75^\circ$.
To determine bulk emission properties, the analysis which follows makes use of 
the formalism presented in~\cite{Weiner2006}, who define a generalised $\chi^{2}$ statistic $L$
 which accounts for errors in both coordinates $e_{x}$ and $e_{y}$, and 
intrinsic scatter $\sigma_{y}$ in a linear model $y=A+Bx$:
\begin{equation}
L=\sum_{i} -\frac{1}{2}\ln2\pi\left(B^{2}e^{2}_{x,i}+e^{2}_{y,i}+\sigma^{2}_{y}\right)+\frac{\left(y_{i}-\left(A+Bx_{i}\right)\right)^{2}}{B^{2}e^{2}_{x,i}+e^{2}_{y,i}+\sigma^{2}_{y}}.
\label{eq:gls}
\end{equation}

$L$ is minimized using standard routines that return errors on the 
fit parameters $A$, $B$ and $\sigma_{y}$; these errors are also verified by using 
bootstrap realizations of the data.
Pixel values are the $x_{i}$ and $y_{i}$, with their errors $e_{x,i}$ and $e_{y,i}$
 determined from the variance maps.
The intrinsic scatter term accounts for the variation in sky signal above that expected
from the instrumental noise alone.
For the purposes of quick reference, a summary of the diffuse analysis is presented
in Table~\ref{tab:difproperties}.

\begin{center}
\begin{deluxetable*}{l|cc|cc}[hb]
\tabletypesize{\scriptsize}
\setlength{\tabcolsep}{0.02in} 
\tablecaption{Average Diffuse Emission Properties \label{tab:difproperties}}
\tablehead{
\colhead{Property} & 
\multicolumn{2}{c}{\underline{\flow} \hspace{2cm}}&
\multicolumn{2}{c}{\underline{\fhigh}} \\
\colhead{} &
\colhead{$\bar{x}$} &
\colhead{$\sigma_{x}$}& 
\colhead{$\bar{x}$} &
\colhead{$\sigma_{x}$} 
}
\startdata
$\alpha_{I} $    & $2.35\pm0.01\pm0.02$  & $0.158\pm0.004$ &    ...         &     ...       \\
                 &                &                 &                &               \\
$P/I$ (\%)       & $1.38\pm0.08\pm0.1$  & $0.74\pm0.03$   & $1.70\pm0.06\pm0.1$  & $1.83\pm0.06$ \\
                 &                &                 &                &               \\
$\phi$ (deg)     & $4.1\pm1.3\pm5$    & $33.7\pm1.0$    & $7.0\pm1.1\pm2$    & $29.7\pm0.9$  \\
\enddata
\tablecomments{Summary of average diffuse emission properties from QUaD data. The symbols
$\bar{x}$ and $\sigma_{x}$ represent the mean and intrinsic scatter of each quantity
in the leftmost column; where quoted, the first error is statistical and the second
systematic.
}
\end{deluxetable*}
\end{center}


\subsection{Total Intensity Spectral Index}
\label{sec:specind}

The spectral index in total intensity is calculated by minimizing Equation~\ref{eq:gls}
to find the slope $B$ (with the intercept $A$ held fixed) between the 
$y_{i}=I_{150,i}$ and $x_{i}=I_{100,i}$ pixels. 
Pixels used in this analysis are restricted to those with $|b|\leq 1^{\circ}$,
where the effects of data processing are smallest --- see Appendix~\ref{subsec:processeffects}.
The slope is simply related to the spectral index as
\begin{equation}
\alpha_{I}= \frac{\ln B}{\ln(150/100)},
\label{eq:specind}
\end{equation}
where the spectral index is calculated at the nominal QUaD center frequencies.
These are assumed fixed, regardless of the spectral index of the underlying 
emission mechanisms --- a more rigorous analysis would involve integrating
each source model over the bandpass and re-calculating the central band frequencies,
as the source spectral index can cause $\nu_{0}$ to shift.
We determined that for reasonable values of the source spectral index, $\nu_{0}$
varies by only a few percent, changing our results insignificantly compared
to absolute calibration errors.
Converting to spectral index via Equation~\ref{eq:specind}, the best fit
parameters are $\alpha_{I}=2.38\pm0.01\pm0.02$, and 
$\sigma_{\alpha{I}}=0.158\pm0.004$.
The first error on $\alpha_{I}$ is statistical, while the second is systematic
as estimated from signal-only simulations (see Appendix~\ref{subsec:specindrecovery}).

\begin{figure}[h]
\resizebox{\columnwidth}{!}{\includegraphics{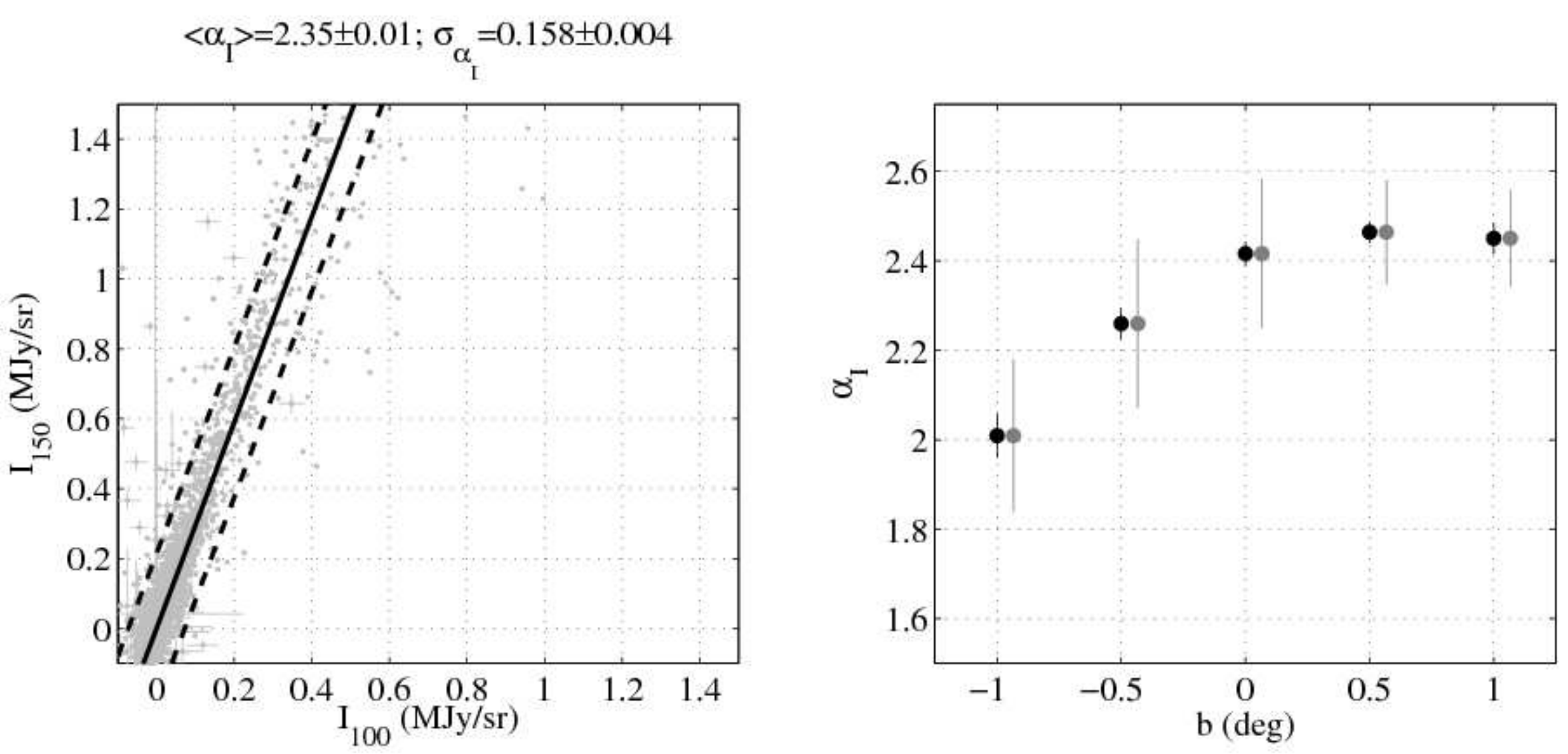}}
\caption{Left: scatter plot of \fhigh~vs \flow~\I\ pixels. The solid black line is
the best fit mean slope $B=I_{150}/I_{100}$, with the broken black lines indicating
the $\pm 1 \sigma_{B}$ regions described by the intrinsic scatter of the data. 
$B$ and $\sigma_{B}$ are converted to spectral index $\alpha_{I}$ and intrinsic 
scatter $\sigma_{\alpha_{I}}$, with the best fit values and errors for these parameters
shown above the plot.
Right: $\alpha_{I}$ as a function of galactic latitude $b$.
The black dots show the mean value and its error, while the gray error bars indicate 
the intrinsic scatter $\sigma_{\alpha_{I}}$.
} 
\label{fig:Ispecind}
\end{figure}

To test for spectral index variations as a function of galactic latitude, the pixels
are subdivided into rows of constant $b$, and the analysis above repeated.
Figure~\ref{fig:Ispecind} shows the results: The spectral index appears to 
flatten between $+1^{\circ}$ and $-1^{\circ}$,
moving from $\alpha_{I}\sim2.4$ to $\alpha_{I}\sim1.8$.
In principle, the field-differencing, filtering and destriping processes
could cause this measurement to be biased due to systematic effects.
However, signal-only simulations described in Appendix~\ref{subsec:specindrecovery}
show that these processes introduce a systematic shift of $<1\%$ and a 
scatter of $\sim7\%$ to the spectral index measurement.
Variation of $\alpha_{I}$ with $b$ in the QUaD data is therefore likely a real
property of the galaxy.

The QUaD value of the spectral index between 100 and \fhigh~is lower than that 
expected for dust alone ($\alpha_{d}\simeq4$), implying one (or more) 
additional emission component is present. 
Possible candidates for the extra component are synchrotron, free-free or spectral
line emission.
Synchrotron has a steeply falling spectral index ($\alpha_{s}\sim-0.7$), but 
is not expected to dominate near the plane on account of its large scale height.
Alternatively, free-free emission has a flatter spectral index 
($\alpha_{ff}\sim-0.1$) and is concentrated towards $b=0$ due to the 
collisional nature of the process.
Given the expected spectral indices of these components, it is likely an
excess of emission in the \flow~band (rather than a \fhigh~deficit) that causes 
the relatively flat QUaD spectral index.
Spectral line emission is a further possibility, with several authors reporting
large line contributions to the bolometric flux, though at higher frequencies
~\citep[e.g.][]{Groesbeck1995,comito2005,wyrowski2006}.
The two QUaD bands alone are insufficient to separate these components; a discussion
of spectral fitting in conjunction with the WMAP data is deferred to 
Section~\ref{subsec:wmapcomp}, while Section~\ref{subsec:sfdcomp} discusses
the relative contributions of synchrotron, dust and free-free as predicted from
models.

\subsection{Polarization}
\label{sec:polfrac}

Two quantites are of interest from the polarization data: the polarization 
fraction, and the angle of polarization.
The analysis of polarization fraction proceeds similarly to the spectral index.
This time, we search for the gradient and intrinsic scatter in a plot of \Q~or \U~
against \I~via minimization of Equation~\ref{eq:gls}; the intercept is fixed 
at zero as before.
Appendix~\ref{subsec:polfracrecovery} demonstrates that field-differencing,
filtering and mapmaking processes bias the recovery of \Qf\ or \Uf\ by 
$\sim0.1\%$.

For the polarization fraction analysis, the jackknife maps are also used to 
test for contamination; the analysis is identical to the signal data.
Note that the polarization fraction for jackknife pixels is defined as
\begin{equation}
\left(\frac{Q}{I}\right)_{jack}=\frac{Q_{jack}}{I_{signal}},
\label{eq:polfracjack}
\end{equation}
and likewise for \U, where $I_{signal}$ is the un-jackknifed total 
intensity map.
Ideally there is no sky signal in the jackknife maps, and hence 
Equation~\ref{eq:polfracjack} represents the polarization fraction that 
would have been observed if the sky had a true polarization fraction 
of zero.

The average \flow~polarization fractions are $Q/I=1.38\pm0.06$\% with $\sigma_{Q/I}=0.54\pm0.02$\%
for the intrinsic variance, and $U/I=-0.12\pm0.05$\% with $\sigma_{U/I}=0.51\pm0.02$\%.
At \fhigh,  $Q/I=1.68\pm0.04$\% with $\sigma_{Q/I}=1.37\pm0.04$\%, 
and $U/I=0.27\pm0.04$\% with $\sigma_{U/I}=1.22\pm0.04$\%.
Combining these results and subtracting the noise bias in $P$ (i.e. 
$P_{debias}=\sqrt{P^{2}-\sigma^{2}_{P}}$), we find average polarization 
fractions of $P/I=1.38\pm0.08\pm0.1$\% at \flow\ and $P/I=1.70\pm0.06\pm0.1$\% at 100
 and \fhigh\ respectively, where the first error is random and the second
due to systematic effects as determined from signal-only simulations
(Appendix~\ref{subsec:polfracrecovery}).
The corresponding intrinsic scatter is simply computed as the quadrature
sum of that from \Qf\ and \Uf, and is $0.74\pm0.03$ and $1.83\pm0.06$ at
100 and \fhigh\ respectively.

\begin{figure}[h]
\resizebox{\columnwidth}{!}{\includegraphics{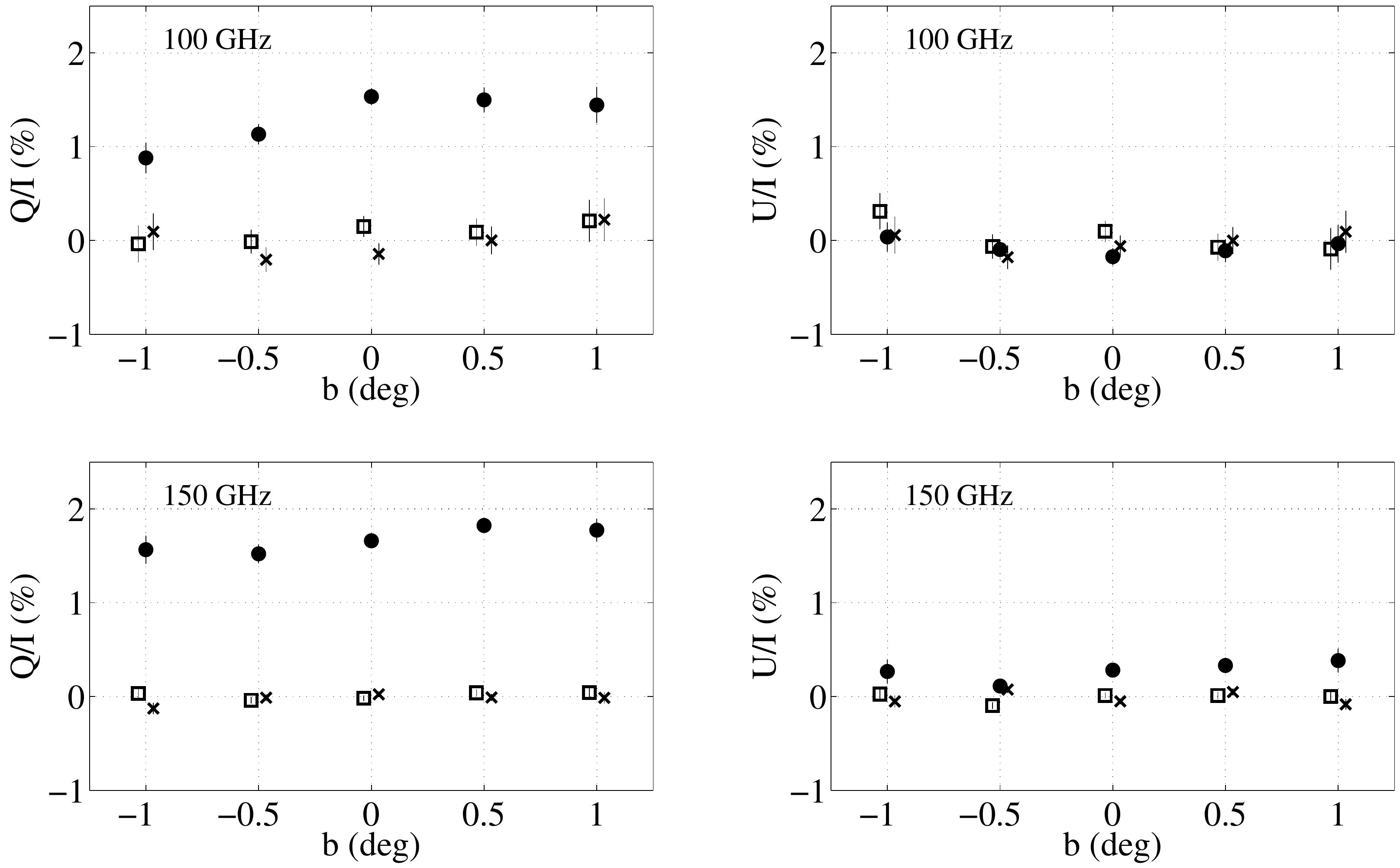}}
\caption{QUaD measurement of galactic-longitude averaged polarization 
fraction as a function of galactic latitude.
Top row is 100 GHz \Qf~(left) and \Uf~(right); the QUaD signal data
are black points, with the scan direction and time jackknife
are squares and crosses respectively (the jackknife data are offset in $b$ from
the signal data for clarity).
Bottom row is for \fhigh.
}
\label{fig:polfrac}
\end{figure}

\vspace{0.01in}
Figure~\ref{fig:polfrac} displays the average \Qf~and \Uf~polarization
 fraction of diffuse emission as a function of galactic latitude for
 the QUaD signal and jackknife data.
The signal data shows polarization fractions which are close to constant
with $b$.
The mean value in each jackknife $b$ bin is expected to be consistent
with zero for both \Qf~and \Uf, with the variance due to pixel noise only --- 
Figure~\ref{fig:polfrac} shows that this is largely the case, with 
the jackknife data in each bin consistent with zero at the $2\sigma$ level or better.

The $P/I$ values of polarization fraction are somewhat lower than the
 Archaeops result~\citep{benoit04}, who found a 4-5\% polarization
fraction for $|b|<2^\circ$ over the galactic longitude range 297 to 85$^\circ$
at 350 GHz.
Conversely, using WMAP 3 year data,~\cite{kogut07} found a 94 GHz polarization 
fraction closer to $\sim1\%$ averaged over a region including the QUaD
survey between galactic latitudes $-10<b<10^\circ$, rising to 3.6\% outside
the P06 mask used in their analysis. 
A direct comparison of QUaD to the WMAP data is presented in 
Section~\ref{subsec:wmapcomp}.

\begin{figure}[h]
\resizebox{\columnwidth}{!}{\includegraphics{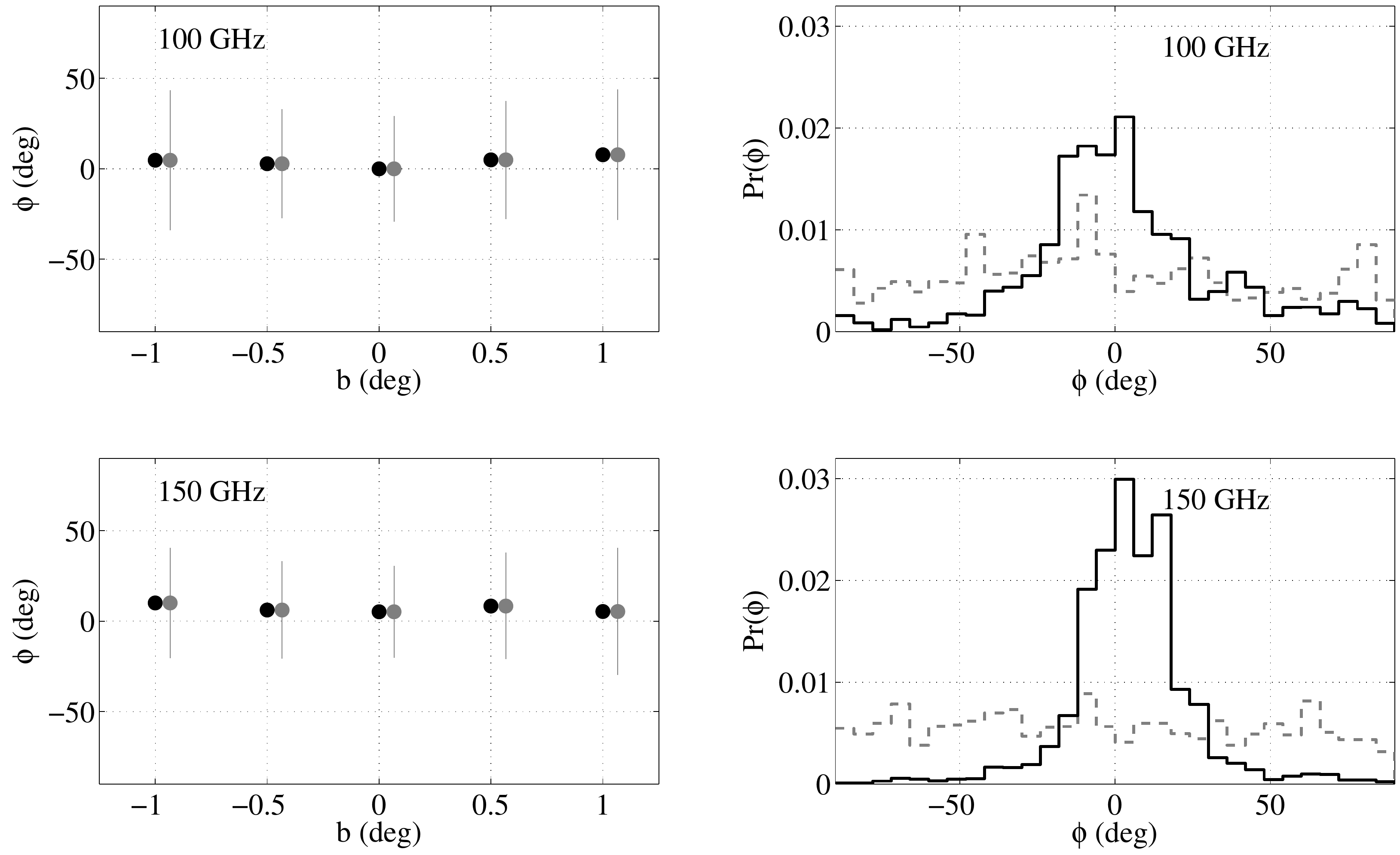}}
\caption{QUaD polarization angle; top and bottom are 100 and \fhigh~respectively.
Left: $\phi$ vs $b$ in 0.5$^\circ$ bins of galactic latitude. 
Black points show the mean $\phi$ in each bin, while black and gray error bars
indicate the statistical error on the mean and the intrinsic scatter respectively.
Right: weighted probability distribution $Pr(\phi)$ of the pixel values as a function of $\phi$. 
The black solid line corresponds to the data, while the broken gray line shows 
Pr($\phi$) for the combined jackknife maps.
The QUaD data clearly show a peak in Pr($\phi$) compared to the jackknife data,
indicating the presence of coherent polarized galactic signal.
}
\label{fig:polangle}
\end{figure}

If the large-scale magnetic field is largely aligned in the plane of
 the galaxy, in the galactic coordinate system this translates to polarized 
emission predominantly in $+Q$.
Figure~\ref{fig:polfrac} demonstrates that this is observed by QUaD, though 
the \Uf~data at \fhigh~shows detected signal.
This observation may be alternatively quantified by directly computing 
the polarization angle $\phi=0.5\mathrm{tan}^{-1}\left(U/Q\right)$ and 
its error $\mathrm{d}\phi=0.5(1+U^{2}/Q^{2})^{-1}\mathrm{d}(U/Q)$, where
\begin{equation}
\mathrm{d}\left(U/Q\right)=\left(U/Q\right)\sqrt{\left(\frac{\mathrm{d}Q}{Q}\right)^{2}+\left(\frac{\mathrm{d}U}{U}\right)^{2}}.
\label{eq:fracerr}
\end{equation} 
\noindent The effective $\chi^{2}$ function in Equation~\ref{eq:gls} is then minimized --- 
this time, since we wish to know the mean and intrinsic variance of the $\phi$
distribution, the gradient term $B$ is fixed at zero.
The analysis is performed on the same pixels as polarization fraction, with the
mean and intrinsic variance calculated for all pixels and in rows of constant
$b$.
Figure~\ref{fig:polangle} shows the results.
The mean polarization angles are $\phi_{100}=4.1\pm1.3^{\circ}\pm5^{\circ}$ and 
$\phi_{150}=7.0\pm1.1^{\circ}\pm2^{\circ}$, with intrinsic variance 
$\sigma_{\phi,100}=33.7\pm1.0^{\circ}$ and $\sigma_{\phi,150}=29.7\pm0.9^{\circ}$.
For $\phi$, the second error is the estimated systematic uncertainty in the polarization
angle using signal-only simulations (see Appendix~\ref{subsec:polfracrecovery}).
In bins of constant galactic latitude, $\phi$ and its intrinsic variance do
 not vary significantly.
A weighted probability distribution of $\phi$ is also shown at each frequency 
for the signal data, and all jackknife data combined. 
At both frequencies, a distinct peak is observed in the distribution close to
the mean values calculated above, while the jackknifes are consistent with
random numbers distributed uniformly between $-90^{\circ}<\phi<90^{\circ}$, as
expected in the presence of no signal.

\begin{figure*}[ht]
\resizebox{\textwidth}{!}{\includegraphics{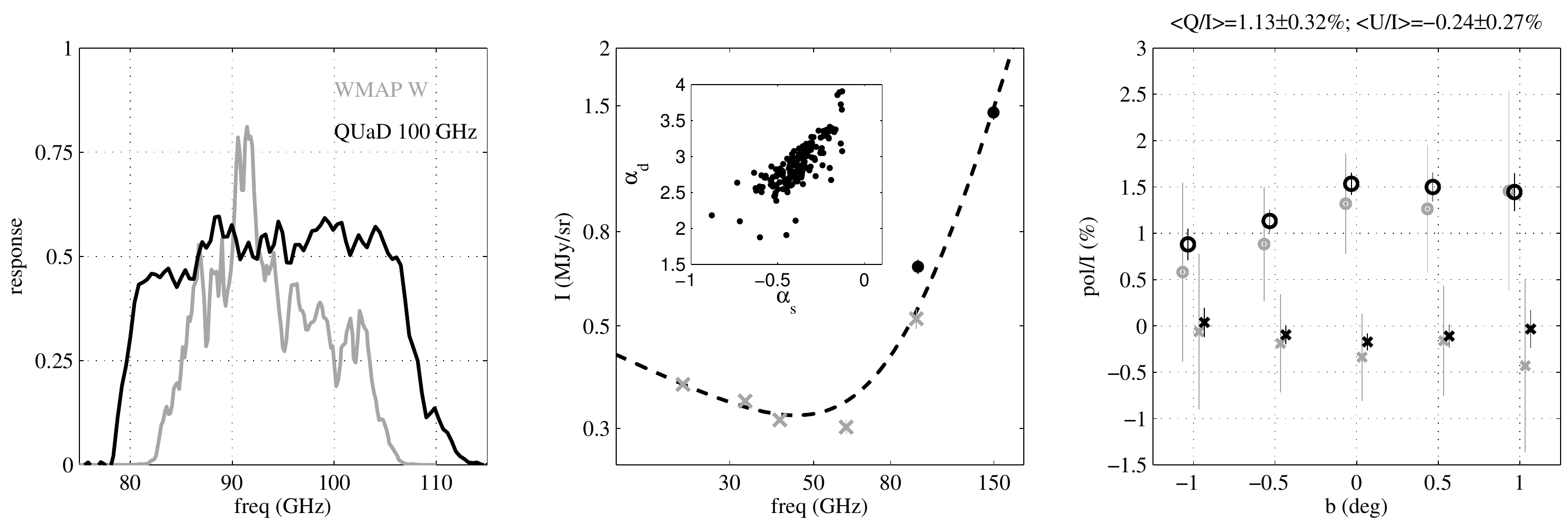}}
\caption{Left: QUaD 100 GHz and WMAP W bandpass spectral response
in black and gray respectively. 
The QUaD band is substantially wider, with $\Delta\nu=27$ GHz compared 
to $\Delta\nu=20$ GHz for WMAP W band. 
Center: Example data points and spectral fit for a representative pixel
 in the WMAP and QUaD maps.
WMAP data are gray crosses, QUaD data are black dots, and the broken black 
line indicates the best-fit two-component spectrum.
The inset shows the distribution of spectral indices taken over map pixels used
in the analysis.
Right: Average polarization fraction as a function 
of galactic latitude.
Gray is WMAP, black is QUaD \flow~for comparison; circles are \Q, crosses
are \U.
The points have been offset in $b$ for clarity.
}
\label{fig:quadvswmap}
\end{figure*}

The QUaD data indicate that while the galactic magnetic field is preferentially
aligned parallel to the plane, there is significant additional scatter present.
However, signal-only simulations (Appendix~\ref{subsec:polfracrecovery}) show that 
a substantial contribution to the scatter in the angle may be present due to 
filtering and processing and effects.
At \flow\ this systematic scatter is comparable to the observed scatter,
indicating we cannot reliably constrain $\sigma_{\phi}$ at this frequency.


\subsection{Comparison to WMAP}
\label{subsec:wmapcomp}

The closest comparison to the QUaD \flow~data is from
WMAP W band, centred on 94 GHz.
For the following analysis, the 5-year WMAP data~\citep{hinshaw2009} are used
to generate simulated timestream in the regions observed by QUaD.
The timestream is processed in the same manner as the signal-only 
simulations described in Appendix~\ref{app:destriping}, in which the 
field-differencing and filtering operations are performed exactly as 
for the QUaD data.

The combination of QUaD and WMAP total intensity data allows 
constraints on the spectral index of emissive components in the galaxy.
Using the QUaD and WMAP beam and pixelization functions, the WMAP Ka, Q, V 
and W band maps, and the QUaD survey, are convolved to the WMAP K-band 
resolution of $0.93^\circ$, and binned into the $0.5^\circ$ pixels used in
the coarse resolution QUaD maps.
Pixel noise in the WMAP maps are determined from regions well away from the
galactic plane, but are much smaller than the absolute calibration uncertainties
of the smoothed maps.
Using the same pixel from each map, the data from each of the seven bands are fit
to a two-component continuum model, the sum of two power laws in frequency $\nu$:
\begin{equation}
I(\nu)=A_{s}\nu^{\alpha_{s}}+A_{d}\nu^{\alpha_{d}}.
\label{eq:specmodel}
\end{equation}
In this expression, $\alpha_{s}$, and $\alpha_{d}$ are the 
synchrotron and dust spectral indices respectively. 
Note that the first component is only loosely termed as due to synchrotron; as
discussed previously, free-free is likely the second most dominant emission mechanism
after dust at 60--100 GHz.
To fit the data, the model is convolved across each bandpass to yield the
average intensity in that band:
\begin{equation}
\tilde{I}_{\nu}=\frac{\int I(\nu)T(\nu)d\nu}{\int T(\nu)d\nu},
\label{eq:specconv}
\end{equation}
where $T(\nu)$ is the bandpass response as a function of frequency, as
shown for WMAP W band and QUaD 100 GHz in the left panel of 
Figure~\ref{fig:quadvswmap}.
The $\chi^{2}$ against the data is calculated and minimized to find the
best-fit parameters for the model in Equation~\ref{eq:specmodel}, with 
Equation~\ref{eq:specconv} evaluated for every set of trial parameters
used in the minimization.

The QUaD data is $\sim25\%$ brighter than WMAP at similar frequencies,
indicating a discrepency between the two experiments which is further
discussed in Section~\ref{subsec:sfdcomp}.
The center panel of Figure~\ref{fig:quadvswmap} shows the fit spectrum for 
a single representative pixel in the data, and a scatter plot of
$\alpha_{s}$ against $\alpha_{d}$ for the spectral fit to each map pixel.
Taken over all pixels, the average spectral indices are 
$\alpha_{s}=-0.32\pm0.03$ and $\alpha_{d}=2.84\pm0.03$.
The former is flatter than might be expected for pure synchrotron 
($\alpha\sim-0.7$), indicating the presence of free-free and/or dust, 
while the latter is lower than that predicted by the FDS dust model 8 
($\alpha_{d,FDS}=3.5$; see Appendix~\ref{subsec:specindrecovery}).
It is clear that the simple two-component model used 
is inadequate to describe the data; this statement remains true if the 
QUaD \flow\ data is excluded.
In fact, \cite{gold08} find that inside the plane (interior to the WMAP 5 year 
KQ95 mask), a ten-parameter model is insufficient to fully describe their 
data, and therefore it may be optimistic to expect simple models to 
account for all emission along lines of sight close to the plane.

The right panel of Figure~\ref{fig:quadvswmap} compares the polarization
fraction, averaged along galactic longitude, between the unsmoothed WMAP W 
band maps, and QUaD \flow --- both maps are binned at $0.5^{\circ}$ resolution.
The plot demonstrates that QUaD is consistent with the WMAP observations to within $1\sigma$,
in both \Q~and \U, with WMAP mean values of $Q/I=1.13\pm0.32\%$ and 
$U/I=-0.24\pm0.27\%$ ($P/I=1.1\pm0.4\%)$, compared to $1.38\pm0.06\%$ and $-0.12\pm0.05\%$
for QUaD ($P/I=1.38\pm0.08\%)$; agreement between the two datasets is also found as a function of
 galactic latitude.

\subsection{Comparison to Emission Models}
\label{subsec:sfdcomp}

Synchrotron and dust models are provided by \cite{finkbeiner01} and \cite{finkbeiner99}
(we use dust model 8 in the latter case); these can be extrapolated into the QUaD bands 
and compared to the observations.
At frequencies $\sim$\flow, free-free emission is also expected to 
contribute.
The H$\alpha$ sky template provided by~\cite{finkbeiner03} is used to
generate a free-free emission map at the QUaD frequencies; we 
follow~\cite{schafer2006}, who convert the H$\alpha$ template from
Rayleigh units into $\mu$K using the formula provided by~\cite{valls1998}:
\begin{equation}
\frac{T_{free-free}(\mu\mathrm{K})}{A_{H\alpha}(R)}\simeq14.0\left(\frac{T_{p}}{10^{4}K}\right)^{0.317}10^{\frac{290}{T_{p}}}g_{ff}\left(\frac{\nu}{\mathrm{10 GHz}}\right)^{-2},
\label{haconv}
\end{equation}
\noindent where $A_{H\alpha}$ are the template pixel values in Rayleighs,
$T_{p}$ is the plasma temperature (assumed to be $10^{4}$K), $\nu$ is the
central observing frequency of the QUaD bands, and $g_{ff}$ is the free-free
Gaunt factor as calculated in~\cite{finkbeiner03}.
The temperature maps are converted to brightness units as in 
Equation~\ref{eq:dbdt}.

Signal-only simulations of synchrotron, dust and free-free are 
used to generate field-differenced and filtered maps of these components,
 which are then summed to produce a model sky at QUaD frequencies.
The limiting factor in resolution is the synchrotron model, which is
derived from 405~MHz maps of~\cite{haslam81,haslam82} with a beam FWHM of
$\sim1^{\circ}$.
QUaD and model maps are binned into $0.5^\circ$ pixels and smoothed to 
$1^{\circ}$ resolution to match the synchrotron model.

\begin{figure}[h]
\resizebox{\columnwidth}{!}{\includegraphics{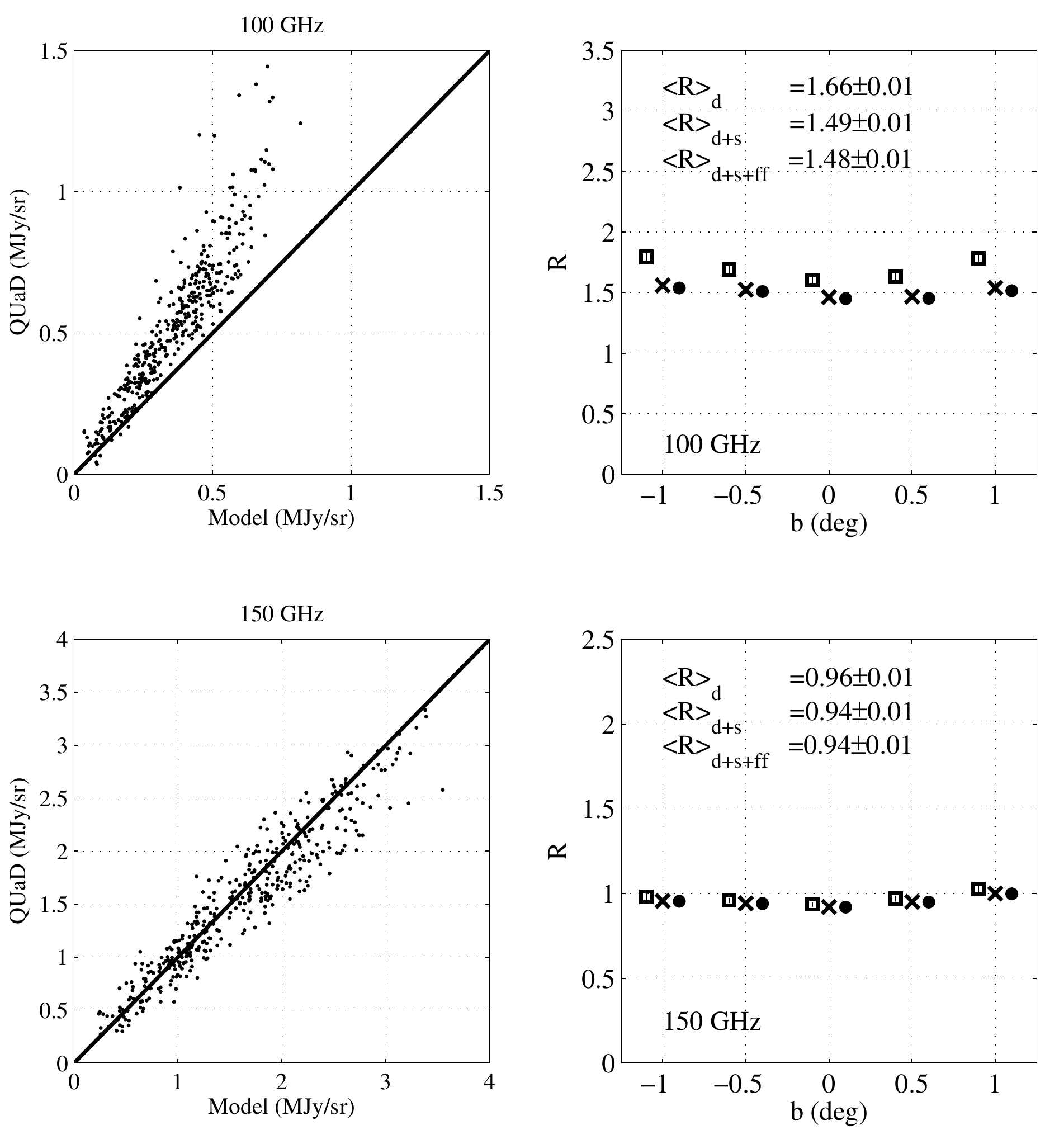}}
\caption{Left panels: QUaD pixel values plotted against signal-only simulations
of the model including dust, synchrotron, and free-free as described in 
the text; top is \flow, bottom is \fhigh.
Right panels: Ratio of QUaD map pixel values to model predictions at 100 (top) and
 \fhigh\ (bottom), as a function of galactic latitude.
Square symbols compare QUaD to the FDS dust model, crosses are dust+synchrotron, and dots 
correspond to dust+synchrotron+free-free.}
\label{fig:quadvssfd}
\end{figure}

In Figure~\ref{fig:quadvssfd} we plot the pixel values between QUaD and
the model predictions against each other (left panels), and the mean 
ratio of pixel values between QUaD and the models $R=I_{QUaD}/I_{model}$ 
in bins of $b$ (right panels).
$R$ is the gradient in a plot of $I_{QUaD}$ against $I_{model}$; this is
 calculated for all pixels and as a function of $b$ by minimizing 
Equation~\ref{eq:gls}, holding the intercept fixed as in 
Section~\ref{sec:specind} with intrinsic scatter fitted simultaneously.
The models are split into three compositions: dust alone, dust+synchrotron,
and dust+synchrotron+free-free, with comparisons to QUaD data made for
each.

At \flow, QUaD is brighter than the FDS dust-only prediction by
a factor $1.66\pm0.01$, decreasing to $1.49\pm0.01$ and $1.48\pm0.01$ as synchrotron
and free-free models are added; the intrinsic scatter is $\sim0.1$ for all
models.
This extra signal at \flow\ is hereafter referred to as the `QUaD excess' --- 
we note that~\cite{gold08} also find an excess of observed W band 
emission over that predicted by the~\cite{finkbeiner99} models over most
of the sky.
At \fhigh~the model predictions are in better agreement with QUaD; $R$ ranges
from $0.962\pm0.007$ (dust only) to $0.942\pm0.006$ (all model components), with
the intrinsic scatter $\sim22\%$.
The ratio $R$ only changes by $1$--$2\%$ comparing QUaD to the model with dust only
or all components at this frequency; this is to be expected since dust 
dominates the emission at \fhigh.
At both frequencies, $R$ does not vary strongly with galactic latitude, 
though care should be taken when interpreting these results since
the maps are smoothed to $1^{\circ}$ resolution, and thus the data points
are highly correlated between $-1^{\circ}<b<1^{\circ}$.


It should be noted a factor $\sim 2$ uncertainty exists in the conversion factor 
from Rayleigh units to antenna temperatures which we apply to the~\cite{finkbeiner03}
free-free maps, which could contribute to the QUaD \flow\ excess.
However, Figure~\ref{fig:quadvssfd} shows that the addition of a free-free
template to a dust+synchrotron model changes the ratio of QUaD to model pixel 
values by less than 1\%; thus even a factor $2$ underestimation of the free-free 
model calibration is insufficient to account for the excess emission.

Another possible explanation of the QUaD excess is molecular line emission: 
large fractional contributions from line emission have 
been measured at higher frequencies towards star-forming 
regions~\cite[e.g.][]{Groesbeck1995,nummelin1998}; these can range from 
$10-65\%$ of the bolometric intensity. 
On the other hand, the COBE FIRAS instrument has detected much lower line emission
contributions ($<$1\%) over large patches of sky~\citep[e.g.,][]{wright91,bennett94,fixsen99}.

If the line emission interpretation is correct, one might expect the QUaD \flow~data to 
be brighter than WMAP W band on account of the wider bandwidth admitting more
lines (see Figure~\ref{fig:quadvswmap}), and the WMAP data to be brighter 
than the models which do not include line emission at all.
This trend is indeed observed: QUaD \flow\ data is a factor $1.48$ brighter
than the model with dust, synchrotron and free-free included, and QUaD \flow\
is also $\sim25\%$ brighter than WMAP W band.
WMAP is therefore some $23\%$ brighter than the combined
models for continuum emission, providing independent evidence that these models
 are insufficient to describe the data close to the plane of the galaxy.

\subsection{Tests of the Spectral Line Hypothesis}
\label{subsec:speclinehyp}

 Given the variation seen in the literature on spectral line contributions at
 higher frequencies, the hypothesis that line emission causes the QUaD excess 
 is subjected to the following additional tests.

\subsubsection{Spectral Comparison Using FIRAS}
\label{subsubsec:speclinefiras}

The FIRAS instrument aboard the COBE satellite provides absolute spectral
measurements covering the QUaD \flow\ band, at a resolution of $\sim7^{\circ}$.
A direct comparison between QUaD and FIRAS is not possible since the FIRAS
beam width is half the maximum width (in R.A.) of the QUaD survey, and hence the QUaD
maps cannot be smoothed to the same spatial resolution.
Neither can spectral discrimination be used to compare the FIRAS data as filtered
 through the QUaD and WMAP bandpasses, since even at maximum spectral
resolution of 3.4 GHz, the frequency sampling is too sparse to resolve the difference
between QUaD and WMAP where there is no spectral overlap.
However, the FIRAS data can be used to obtain an upper limit on the emission 
expected in the QUaD \flow\ band, providing a useful consistency check of the QUaD absolute
calibration.
We restrict ourselves to pixels less that $5^{\circ}$ from the galactic plane, and
subtract the best-fit CMB monopole blackbody spectrum from each FIRAS frequency 
channel. 
Other filtering effects such as polynomial subtraction are not included because
the large FIRAS beam smooths the galactic signal to greater galactic latitudes than
covered in the QUaD survey.
Polynomial subtraction of the signal lying at the QUaD scan ends would then
reduce the amplitude below the level of signal loss due to filtering in the QUaD
survey itself.

The signal-to-noise at $\sim$\flow\ is low in the FIRAS data, so an upper limit
is obtained by finding the 95th percentile of the pixel distribution for each 
frequency channel.
This `spectrum' of 95th percentiles is then integrated over the normalized QUaD 
\flow\ bandpass, resulting in an upper limit of 5.3 MJy/sr at 95\% confidence.
This limit is consistent with the QUaD data; Figure~\ref{fig:brightmaps} shows that
at \flow\ the peak brightness is $\sim7 \mathrm{MJy/sr}$ near the galactic
center, which would be reduced by the ratio of beam areas (a factor 
$\sim(7\times60/5)^2=7056$) once beam smoothing is taken into account.
Therefore the FIRAS data cannot isolate the QUaD excess as being due to spectral
lines or an absolute calibration mismatch.

\subsubsection{Spectral Comparison Using CO 1-0 Transition Maps}
\label{subsubsec:speclineco}

The most prominent line near the lower QUaD band is the 1--0 CO transition at $\sim115$ GHz.
This has been mapped over the inner galactic plane by~\cite{dame01} and references
therein; the survey has spatial resolution $6'$, very close to QUaD at \flow. 
As seen in Figure~\ref{fig:quadvswmap}, the QUaD band response at 115 GHz is small (a
factor of $\sim1000$ smaller than the peak in fact).
The CO maps are restricted to the QUaD survey boundaries, and the antenna temperature
units converted to MJy/sr, first by converting antenna temperatures to thermal temperature
units for each frequency channel, and then by averaging the CO data in frequency over the 
spectral bandwidth, and finally averaging over the QUaD bandpass.
We find a peak CO contribution of $10^{-2}~\mathrm{MJy/sr}$, or approximately 2\% of the typical
brightness of a WMAP pixel in the QUaD survey region, and is therefore insufficient to 
account for the $\sim25\%$ excess seen by QUaD over WMAP.

\subsubsection{Spectral Line Check Using Polarization}
\label{subsubsec:speclinepol}

Spectral lines are not expected to emit polarized radiation, so any contribution
can be tested by comparing the unpolarized and polarized data.

\begin{figure}[h]
\resizebox{\columnwidth}{!}{\includegraphics{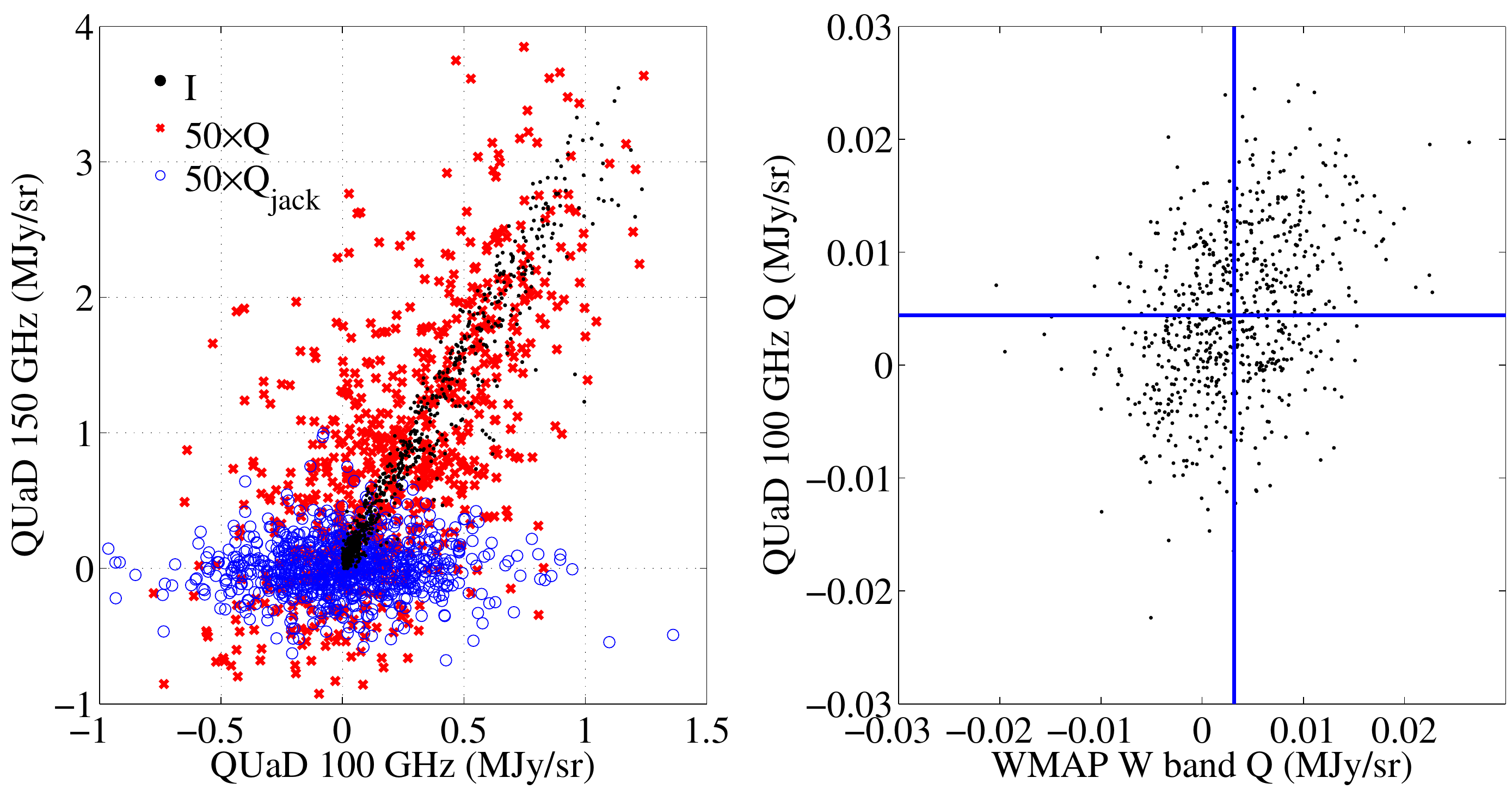}}
\caption{Left: Scatter plot of QUaD 150 vs. \flow\ data.
Black points are for unpolarized intensity \I, red are for \Q,
and blue are for the scan-direction \Q\ jackknife to indicate
the level of noise present in the QUaD \flow\ \Q\ data.
The polarization data have been multiplied by 50 for visual display
purposes only.
Right: Scatter plot for QUaD \flow\ data against WMAP W band for
Stokes \Q.
The solid lines indicate the mean value of each data set.}
\label{fig:reffig}
\end{figure}

The left panel of Figure~\ref{fig:reffig} shows a plot of QUaD 150 vs. \flow\ 
data in \I\ and \Q.
The slope in \I\ is clear, and is apparently traced by \Q; this indicates
that line emission does not contribute significantly to the lower QUaD band, as 
otherwise a steeper slope would be expected in polarization.
However, the signal-to-noise at \flow\ is low: Also plotted in 
Figure~\ref{fig:reffig} are the pixel values from the scan-direction
jackknife, which demonstrate that instrumental noise contributes significantly to
the \flow\ data, and therefore statistical uncertainties could bias the polarized
 spectral index measurement towards lower values.

Greater signal-to-noise can be acheived by using the galactic longitude-averaged data.
Neither dust nor synchrotron are expected to vary polarization fraction with
frequency, yet as seen in Table~\ref{tab:difproperties} and Figure~\ref{fig:reffig}, 
the observed polarization fractions differ between QUaD bands.
Taking the average values of $P/I$ at each frequency, the unpolarized spectral
index, and combining the measurement uncertainties and intrinsic scatter into a single
error term, we find $\alpha_{P}=2.81^{+0.26}_{-0.25}$.
This is discrepent with the total intensity spectral index at the $\sim2\sigma$
level, and is thus an inconclusive test on the existence or not of additional 
emission components.

As a final test, the QUaD \flow\ maps are compared to WMAP W band in Stokes \Q\ 
on a pixel-by-pixel basis.
The right panel of Figure~\ref{fig:reffig} shows such a plot of this test; 
both data sets are noisy and a statistically significant measurements of the gradient
is not possible.
Instead, we simply take the mean of each set of pixels and calculate the ratio,
finding $\langle Q\rangle_{QUaD}/\langle Q\rangle_{WMAP}=1.4\pm0.5$.
Such large errors render the data insufficient to measure an excess of QUaD 
polarization over WMAP, and thus cannot verify or falsify a contribution due to
line emission or an absolute calibration mismatch.

\section{Conclusions}
\label{sec:conclusions}

We present the QUaD survey of the Milky Way in Stokes $I$, \Q~and \U, with 
a resolution of 5 (3.5) arcmin at $100$ and \fhigh\ respectively.
The survey covers two regions, $110^{\circ}<\mathrm{RA}<190^{\circ}$ and 
$210^{\circ}<\mathrm{RA}<290^{\circ}$, both in the dec range 
$-60.5^{\circ}<\mathrm{dec}<-26.5^{\circ}$, corresponding to approximately 
$245-295^\circ$ and $315-5^\circ$ in galactic longitude, and 
$-4<b<4^\circ$ in galactic latitude --- a total of $\sim800$ square degrees.

Degrading the map resolution to $0.5^\circ$ pixels, the average spectral index of 
diffuse emission is $\alpha_{I}=2.35\pm0.01\mathrm{(stat)}\pm0.02\mathrm{(sys)}$, 
assuming nominal band centers of 94.5 and 149.6 GHz.
This value of spectral index is flatter than that expected from
dust alone --- \cite{gold08} demonstrate that the WMAP 5 year data only weakly
constrains the dust spectral index, but use a prior range $3.5<\alpha_{d}<5$, indicative
of the expectation for this component.
The low QUaD-only value is interpreted as evidence for additional emission 
components in the lower frequency QUaD band.

A direct comparison to WMAP 5 year W band data shows the QUaD \flow\ maps are on 
average $\sim25\%$ brighter.
Fitting a two-component continuum model to all pixels in the five WMAP and two QUaD 
bands results in constraints of $\alpha_{s}=-0.32\pm0.03$ and $\alpha_{d}=2.84\pm0.03$.
The first is attributed to a composite of synchrotron, free-free and
dust expected at $\sim100$ GHz close to the galactic plane, with the 
second interpreted as mostly dust.
However, the fit is poor for such a simple model and more emission components
would be required to fully explain the data.

Similarly to the spectral index determined from QUaD alone, $\alpha_{d}$ is
lower than the expectation from available models.
A composite model of dust, synchrotron and free-free emission underestimates 
the brightness at \flow, where QUaD observes on average a factor $1.48\pm0.01$ more 
signal.
One interpretation of this effect is molecular line emission.
This possibility is discussed further in Section~\ref{subsec:speclinehyp}, where a variety
of cross-checks indicate that the 115~GHz CO 1-0 line is unlikely to be the main cause
of the \flow\ excess, but that the QUaD data are consistent with absolute spectral 
measurements from the COBE FIRAS instrument.
At \fhigh~ the agreement is better, with an average pixel ratio of 
$0.942\pm0.006$ between QUaD and the models. 

The QUaD data allow measurement of the polarization fraction in both bands --- the
results quoted here are taken from galactic coordinates maps using the IAU 
convention for Stokes parameters $Q$ and $U$.
Analysis in the Source Paper shows that few compact objects have 
measurable polarization and thus we assume the dominant source of the polarized 
emission studied here is diffuse.
On average, $Q/I=1.38\pm0.06$\% and $1.68\pm0.04$\% at 100 and 150 GHz, with
the equivalent averages for \Uf~being $-0.12\pm0.05$\% and $0.27\pm0.04$\%. 
The intrinsic scatter on these quantities were found to be typically $\sim0.5$ 
($\sim1.3$)\% at 100 and \fhigh\ respectively, reflecting fluctuations in 
polarization fractions at different positions in the galactic plane.
Signal-only simulations indicate that the systematic error on polarization
fraction is of order $0.1\%$ at both frequencies.
Measurements of \Qf\ and \Uf\ are also possible as a function of 
galactic latitude $b$ within $|b|\leq1^\circ$, and show evidence 
of small deviations from the average polarization fractions quoted above, 
but within the range allowed by the intrinsic scatter.
Combining \Qf\ and \Uf, we find total polarization fractions $P/I=1.38\pm0.08\pm0.1$\% 
at \flow\ and $1.70\pm0.06\pm0.1$\% at \fhigh, where the first error is random
and the second systematic.
The intrinsic scatter at these frequencies is $0.74\pm0.03$ and $1.83\pm0.06$.

Comparing the QUaD polarization fraction to that from WMAP 5 year data,
 agreement is found between the two datasets at QUaD \flow\ and
 WMAP W band, the latter giving an average polarization fraction $1.1\pm0.4$\%.
The polarization fraction measurements reported here provide encouragement that 
large areas of sky may be useful for probing inflationary cosmology with CMB
 polarization B-modes at frequencies above 100 GHz.

The mean angle of polarization close to the plane measured by QUaD is 
is $<\phi>=4.1\pm1.3\pm5^{\circ}$ at \flow, with $<\phi>=7.0\pm1.1\pm2^{\circ}$ 
at \fhigh, where the first error is random and the second systematic.
Although the data indicate QUaD has detected intrinsic scatter in the distribution
of $\phi$, the amount of scatter is consistent with that introduced by
filtering and map processing effects. 
The observations therefore provide evidence of the large-scale alignment of the galactic
magnetic field.

Extensive tests could not definitively isolate the cause of the QUaD excess
signal observed over WMAP W band and continuum emission models at \flow.
The QUaD data is consistent with an unpolarized intensity upper limit derived 
from FIRAS data, ruling out a large QUaD absolute calibration error.
A second test of the absolute calibration is calculated from the ratio of
 average Stokes \Q\ in QUaD and WMAP pixels; we find 
$\langle Q\rangle_{QUaD}/\langle Q\rangle_{WMAP}=1.4\pm0.5$, an excess over \I\ 
on average, but statistically consistent with the mean total intensity 
excess of $\sim1.25$.
Low signal-to-noise in polarization at \flow\ prevented a direct determination
of the polarized spectral index; a larger value than total intensity would be
evidence for line emission in the \flow\ band.
Using a combination of the average polarization fraction and the unpolarized 
spectral index, we determined $\alpha_{P}=2.81^{+0.26}_{-0.25}$, within $2\sigma$
of $\alpha_{I}$ and thus inconclusive regarding molecular line emission.
Maps of the CO 1--0 transition~\citep{dame01} multiplied through the QUaD bandpass
placed an upper limit on the contribution of this molecular line of 
$10^{-2}~\mathrm{MJy/sr}$, which at $\sim2\%$ of the typical brightness of a
WMAP W band pixel is insufficient to account for the QUaD excess.
We conclude that higher signal-to-noise measurements of the polarized galactic
emission at \flow\ are required to resolve the QUaD excess, and should be provided
in the near future by the Planck satellite~\citep{Planck}.

This paper focussed on the properties of diffuse emission (typically $0.5^{\circ}$ 
scales and larger); however, the QUaD galactic plane survey contains
 information on scales down to 5 and 3.5 arcminutes at 100 and \fhigh~.
The small-scale properties of emission via discrete sources are explored
in the Source Paper, where a variety of objects have been detected, such
as ultra-compact HII regions and supernova remnants.

\section*{Acknowledgements}

This paper is dedicated to the memory of Andrew Lange, who gave wisdom 
and guidance to so many members of the astrophysics and cosmology 
community. 
His presence is sorely missed.
We thank our colleagues on the BICEP experiment and Dan Marrone 
for useful discussions.
QUaD is funded by the National Science Foundation in the USA, through
grants ANT-0338138, ANT-0338335 \& ANT-0338238, by the Science and
Technology Facilities Council (STFC) in the UK and by the Science
Foundation Ireland.  The BOOMERanG collaboration kindly allowed the
use of their CMB maps for our calibration purposes.  MZ acknowledges
support from a NASA Postdoctoral Fellowship.  PGC acknowledges funding
from the Portuguese FCT. SEC acknowledges support from a Stanford
Terman Fellowship. JRH acknowledges the support of an NSF Graduate
Research Fellowship, a Stanford Graduate Fellowship and a NASA
Postdoctoral Fellowship. YM acknowledges support from a SUPA Prize
studentship. CP acknowledges partial support from the Kavli Institute
for Cosmological Physics through the grant NSF PHY-0114422.  EYW
acknowledges receipt of an NDSEG fellowship.
We acknowledge the use of the Legacy Archive for Microwave Background 
Data Analysis (LAMBDA). Support for LAMBDA is provided by the NASA 
Office of Space Science.

\bibliographystyle{apj}
\bibliography{ms}

%
\begin{appendix}

\section{Further Details on Destriping Algorithm}
\label{app:destriping}

\subsection{Map Destriping}
\label{subsec:destripemaps}

When constructing the initial maps $m_{0}$ and $m_{1}$, the choice 
of polynomial filtering order is a trade-off between increased atmospheric noise
reduction (higher order) and reduced filtering of the galactic signal of interest 
(lower order).
A simple DC-level+slope filter function is used in the QUaD survey because
we are interested in the emission properties on both small and large
angular scales, both of which are suppressed or corrupted by a higher order
filter function.
However, this choice results in a larger atmospheric noise contribution
than would be present had a higher order polynomial been used --- 
the $m_{0}$ map in Figure~\ref{fig:quadfiltstages} exhibits large
row-to-row striping as a result, which we suppress as follows.
After filtering, $1/f$ noise is largely uncorrelated
 between rows of pixels, while galactic structure is strongly 
correlated on these angular scales due to its intrinsic structure and 
beam smoothing.
Smoothing the $m_{1}$ maps with a circularly symmetric gaussian kernel 
of width $4\times\sigma_{beam}$ at each frequency reduces striping 
between pixels; the smoothed map is treated as a template map of the sky, $m_{t}$.
From $m_{t}$, `signal only' timestream $d_{t}$ is interpolated
 and subtracted from the original data $d$:
\begin{equation}
d_{n}=d-d_{t}.
\label{eq:noisetod}
\end{equation}
The `signal-subtracted' timestream $d_{n}$ is now dominated by 
atmospheric noise with little galactic signal present, and is fit with a
higher order polynomial to measure the atmospheric modes (a 6th-order
polynomial is used in the QUaD survey).

\begin{figure}
\resizebox{\columnwidth}{!}{\includegraphics{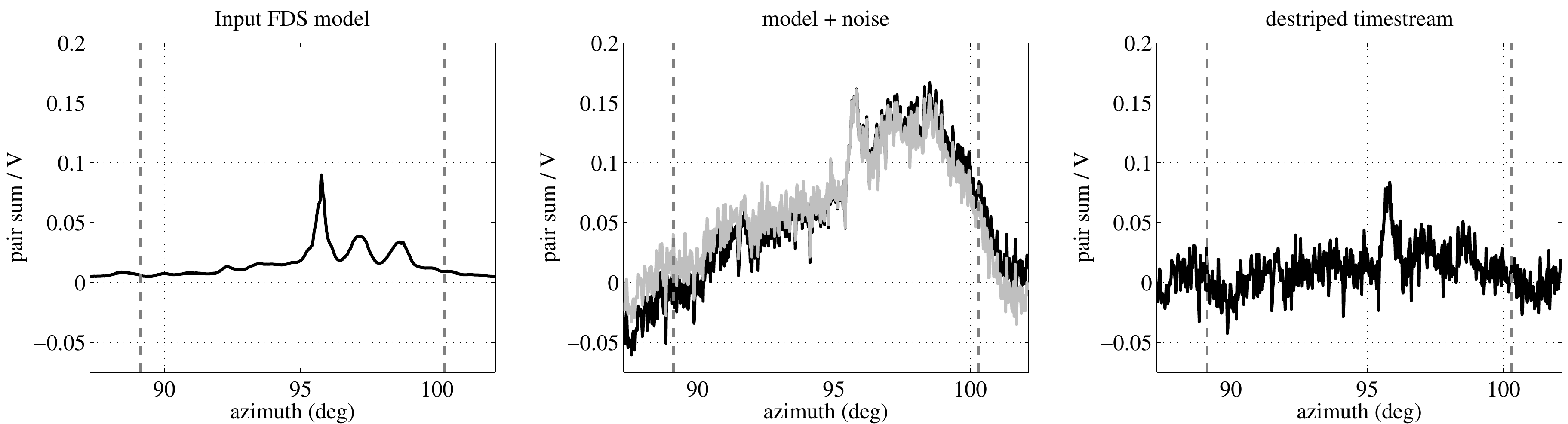}}
\caption{Destriping process for one half-scan of data for the central \fhigh~
pixel. 
The dashed graylines show the regions of the scan ends used for DC-level 
plus slope filtering. 
Left: Simulated input FDS model 8 signal. 
Center: Input model plus simulated noise before (black) and after 
(gray) initial filtering. 
Right: Destriped timestream --- the recovered timestream is 
an improved estimate of the true sky signal compared to the initial
filtering.}
\label{fig:destripetod}
\end{figure}

Using the resulting polynomial coefficients $p_{n}$, a polynomial $d_{p_{n}}$
which represents atmospheric modes is subtracted from the original 
unfiltered timestream
\begin{equation}
d_{s}=d-d_{p_{n}},
\label{eq:sigtod}
\end{equation}
and the filtered data is coadded into the map.
Since $d_{p_{n}}$ largely measures atmospheric $1/f$ modes, the noise in 
maps made from $d_{s}$ (denoted $m_{2}$ in the map ordering of our algorithm) is
considerably whiter than the $m_{0}$ maps.
The improvement may be seen in simulated timestream in Figure~\ref{fig:destripetod},
and in the resulting maps in Figure~\ref{fig:filtstages}; the procedure is 
repeated independently for \I, \Q~and \U~maps at both frequencies.

Some subtleties are present in this method; since $m_{t}$ is a 
smoothed version of the sky as observed by QUaD, subtracting it
from the timestream introduces residuals near the locations of
bright point sources (effectively, subtracting two gaussian functions
of equal cross-sectional area but differing widths).
These residuals can be large, particularly near the galactic center,
influencing the $p_{n}$, causing spurious filtering residuals in the $m_{2}$
maps.
This effect is reduced by locating bright sources in $m_{1}$ as described
in the Source Paper and then masking them as described below.
Since $1/f$ noise in the $m_{1}$ total intensity maps
 can be spuriously detected as false sources, a high signal-to-noise 
threshold of $S/N>10$ is used in the \I~map.
For \Q~and \U~the noise is considerably whiter, so a threshold of 
$S/N>3$ is used.

With the locations of bright sources known, two methods are used to 
reduce their impact on higher-order filtering.
First, the source pixels are replaced with a local median and the
smoothed template map $m_{t}$ is computed --- this reduces the amount
of source power smeared out by the smoothing, lowering the corresponding
residuals in $d_{n}$.
Second, the source locations are masked with a conservative radius of
$6\sigma_{beam}$ when performing the 6th-order polynomial fit.

After these steps, the data is filtered as in Equation~\ref{eq:sigtod} and coadded
into the map, with post-filter inverse scan variances used as weights.
Note that the variances for the destriped data are computed over the
entire scan after template subtraction and filtering, and with point 
sources masked, i.e $\sigma^{2}=var(d_{n}-d_{p_{n}})$.

The mapping algorithm implemented in this paper can be
summarized as follows:

\newcounter{step}
\begin{list}{\bfseries\upshape Step \arabic{step}:}
  {\usecounter{step}
    \setlength{\labelwidth}{2cm}\setlength{\leftmargin}{1.5cm}
    \setlength{\labelsep}{0.5cm}\setlength{\rightmargin}{1cm}
    \setlength{\parsep}{0.5ex plus0.2ex minus0.1ex}
    \setlength{\itemsep}{0ex plus0.2ex} \slshape}
\item Construct $m_{0}$ map from timestream filtered with DC-level+slope 
determined from scan ends.
\item Locate sources using the method described in the Source Paper, 
  mask sources and and repeat Step 1 above to give $m_{1}$.
\item Smooth $m_{1}$ with gaussian kernel of 
$\sigma_{t}=4\times\sigma_{beam}$, with bright source pixels
replaced by local median - this is the template map $m_{t}$.
\item Interpolate `signal-only' timestream $d_{t}$ from $m_{t}$, 
and subtract from the original data $d$ to give an approximate
 `noise-only' timestream $d_{n}$.
\item Fit $d_{n}$ with high order polynomial, masking sources
 located in Step 2. This step measures the `noise-only' polynomial 
  coefficients $p_{n}$.
\item Filter the original timestream $d$ using a polynomial 
with coefficients $p_{n}$, and coadd the filtered half-scans 
into the map using post-filter inverse half-scan variances as 
weights; this produces the final $m_{2}$ maps.
\end{list}  

Though the destriping algorithm is implemented entirely in
timestream/image space, it has a direct interpretation as 
a linear filter in Fourier space as discussed in 
Section~\ref{subsec:fpdestriping}.

\subsection{Test of Mapping Algorithm}
\label{subsec:destripetests}

\begin{figure}[ht]
\resizebox{\columnwidth}{!}{\includegraphics{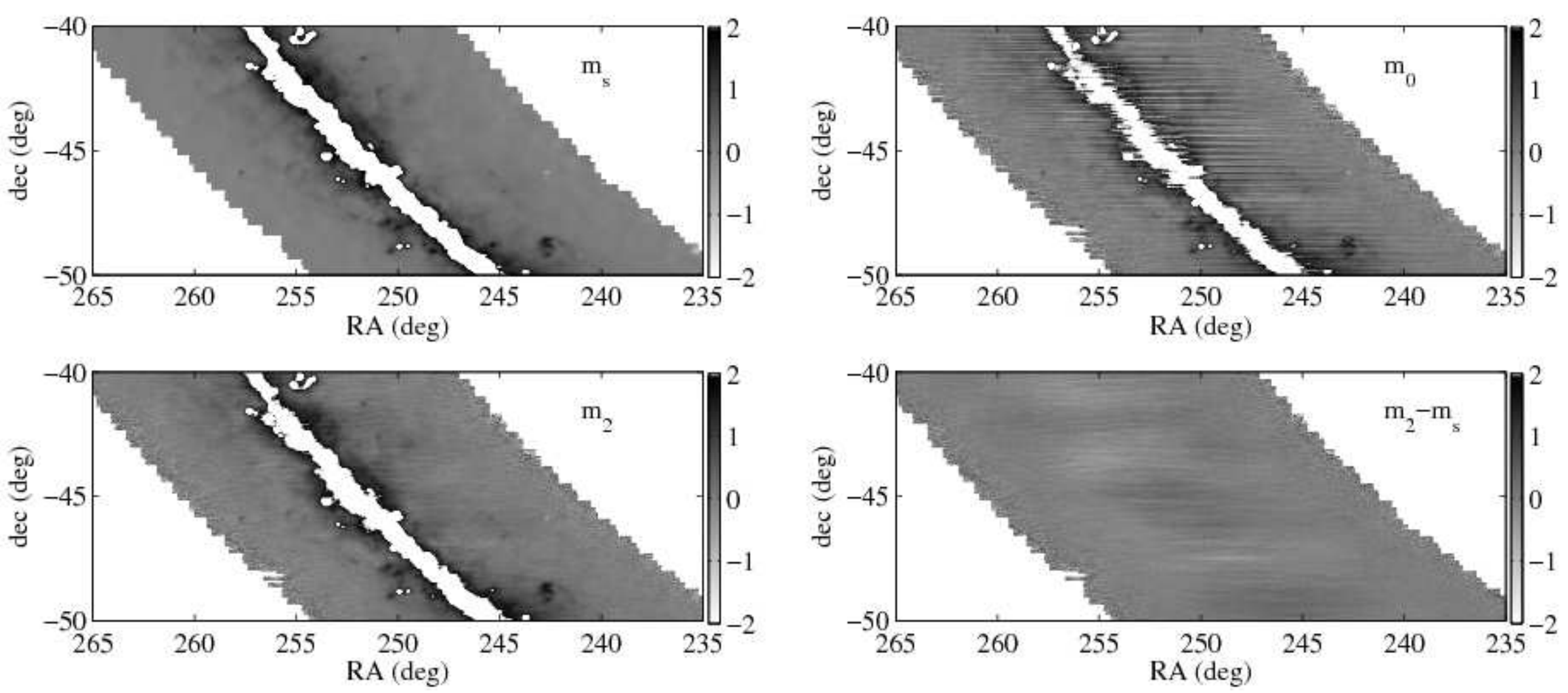}}
\caption{Maps illustrating the major stages of the mapmaking
process. All colorscales are in MJy/sr, with saturated pixels within 
the survey region shown in white.
{\it Top left:} Signal-only FDS model 8 prediction at \fhigh.
{\it Top right:} signal plus noise simulation of the same model
after DC-level + slope filtering, i.e. $m_{0}$ stage map.
{\it Bottom left:} Recovered $m_{2}$ map after destriping.
{\it Bottom right:} Difference between input and destriped maps. 
The color stretch has been reduced by a factor 5 to enhance the residuals;
the noise consists mainly of white noise, and large-scale modes which are
removed before the destriping in Step 4 of Section~\ref{subsec:destripemaps}.}
\label{fig:filtstages}
\end{figure}

To test the algorithm described above, signal-only simulations
of the FDS dust model 8 evaluated at the QUaD center frequencies 
are generated as mock sky signal.
We wish to add realistic noise to the simulations, but since the 
galactic signal can dominate the timestream, we cannot 
directly take the power spectrum of the data and use 
it to regenerate noise as in P09 --- doing so results in 
noise strongly correlated with regions of bright galactic 
emission.
To mitigate this effect, the sky signal (as estimated from the destriped maps) is 
first removed from the data, and the resulting sky-subtracted timestream 
processed  in the same way as P09: continuous segments of data are Fourier
transformed, binned, and the covariance matrix of the fourier
 modes taken between all channels in each bin separately.
Noise timestream is then regenerated by mixing uncorrelated random 
numbers with the Cholesky decomposition of this covariance matrix.
Taking the inverse Fourier transform yields simulated noise timestream
 with the observed degree of covariance as the signal-subtracted data.
P09 demonstrates that this process yields simulated noise which 
is indistinguishable from the real.
The simulated noise is added to the signal-only timestream, coadded 
into $0.02^{\circ}$ maps, and destriped according to Section~\ref{subsec:destripemaps}.
Figure~\ref{fig:filtstages} shows the signal-only input map
$m_{s}$, the DC+slope filtered map $m_{0}$, the destriped map
$m_{2}$, and the residual map $m_{2}-m_{s}$.
Qualitatively, it is clear that the algorithm is effective at suppressing $1/f$
noise, and introduces small residuals compared to the galactic 
signal of interest.

\subsection{Effects of Map/Timestream Processing}
\label{subsec:processeffects}

The mapmaking algorithm described above results in a loss of 
signal due to the main stages of processing: field-differencing 
and destriping.
To test the effects of each, the same signal-only simulations
 of FDS model 8 as in Section~\ref{subsec:destripetests} are used, comparing
pixel values at each (cumulative) stage of processing of the input
 maps.

\begin{figure}[h]
\resizebox{\columnwidth}{!}{\includegraphics{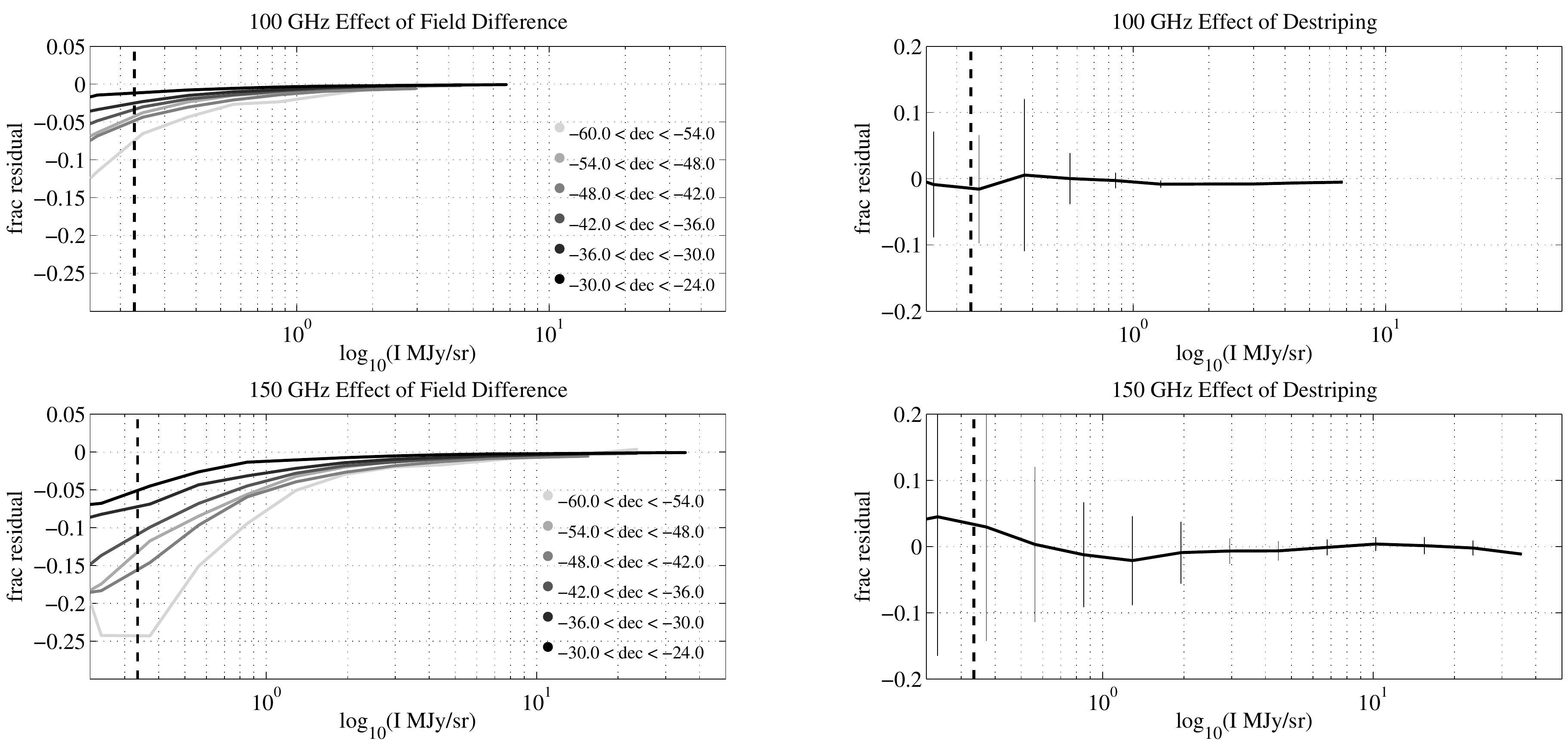}}
\caption{Effect of field differencing and destriping on map pixel values,
for FDS model 8 signal-only simulations.
Top and bottom are \flow~and \fhigh~respectively.
Left: Mean fractional change in pixel values of $m_{0}$ map, before and 
after field differencing, as a function of input pixel value. 
Lines in different shades of gray correspond to different bins in dec, as denoted
in the legend. 
Right: Mean fractional change in pixel values before and after destriping
process (i.e. comparing $m_{1}$ and $m_{2}$ maps).
The error bars correspond to the standard deviation in each bin in 
$\log_{10} I~ \mathrm{(MJy/sr)}$.}
\label{fig:filterfx}
\end{figure}

The left panels of Figure~\ref{fig:filterfx} show the median fractional
change in pixel values of $m_{0}$ maps, before and after 
field-differencing.
At the mean $3\sigma$ noise level in the QUaD maps, 
field-differencing reduces the signal by $1-7\%$ at \flow, and
$5-25\%$ at \fhigh, depending on dec.
The declination dependence arises from the fact that our $15^\circ$
azimuth scans correspond to a smaller RA range at lower dec, scaling as
$\mathrm{cos}(\mathrm{dec})$.
Therefore the low dec trail field scans lie closer to the galactic plane, 
and so subtract out more sky signal when the lead and trail fields are differenced.
The fractional loss in signal is smaller for brighter pixels close
to the galactic plane, with less than $5\%$ reduction above 0.35 (1.3)
MJy/sr at 100 (150) GHz over all dec.

A similar analysis is shown for the effect of destriping in the right 
hand panels of Figure~\ref{fig:filterfx}.
Since the destriping method only reduces uncorrelated noise between rows
of pixels, it is not dec-dependent.
Destriping does introduce fluctuations in the pixel values, shown by the
error bars in Figure~\ref{fig:filterfx}.
The fractional rms (relative to the signal) is at most $15\%$ for pixels
 with amplitude above the $3\sigma$ noise level at both frequencies, and
decreases with increasing signal amplitude --- the rms is $<5\%$ above
0.5 (0.8) MJy/sr at 100 (150) GHz.

\subsection{Fourier Plane Interpretation of Map Destriping Algorithm}
\label{subsec:fpdestriping}

\begin{figure*}[ht]
\resizebox{\textwidth}{!}{
\includegraphics[width=5.5in]{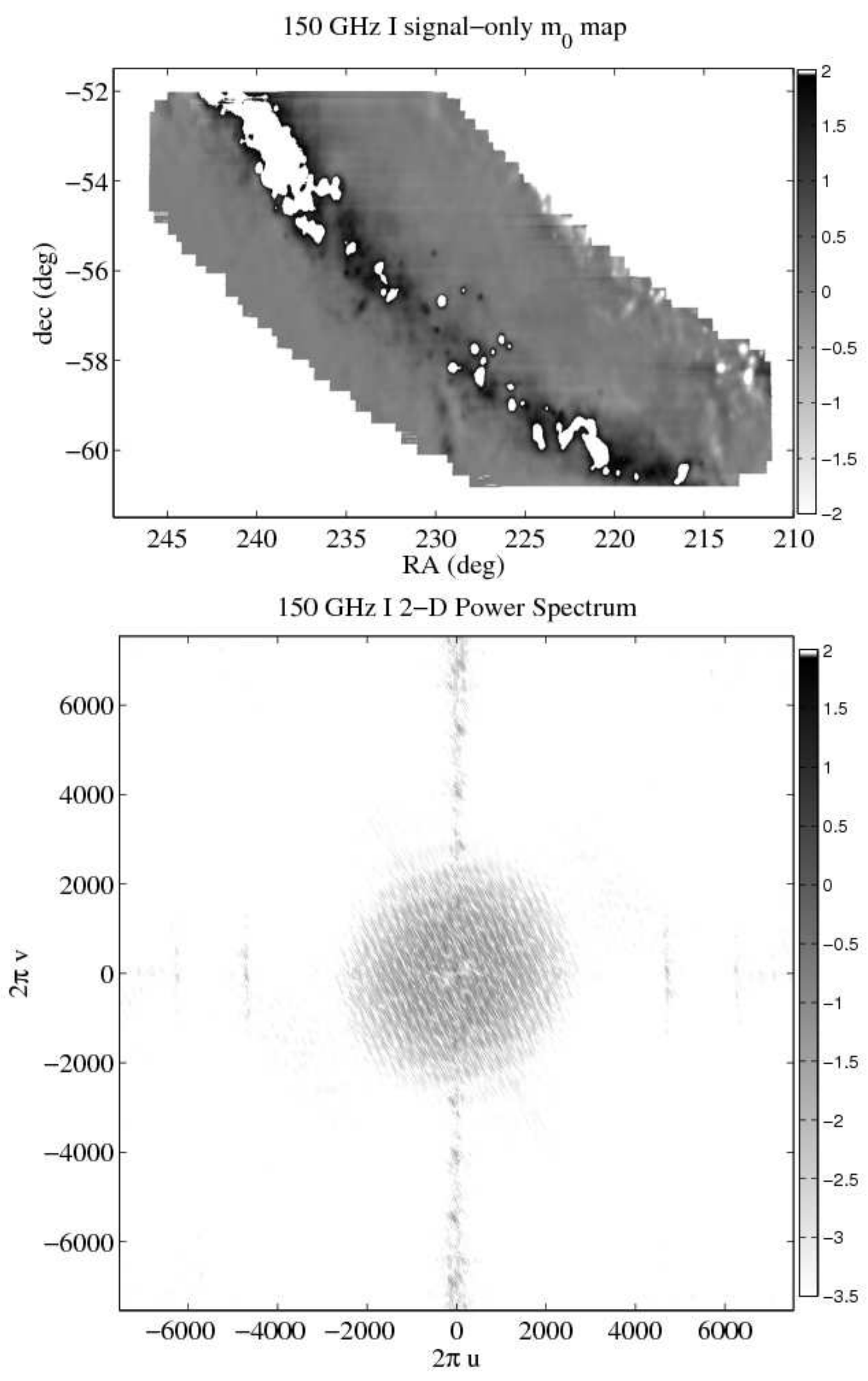}   \\
\includegraphics[width=5.5in]{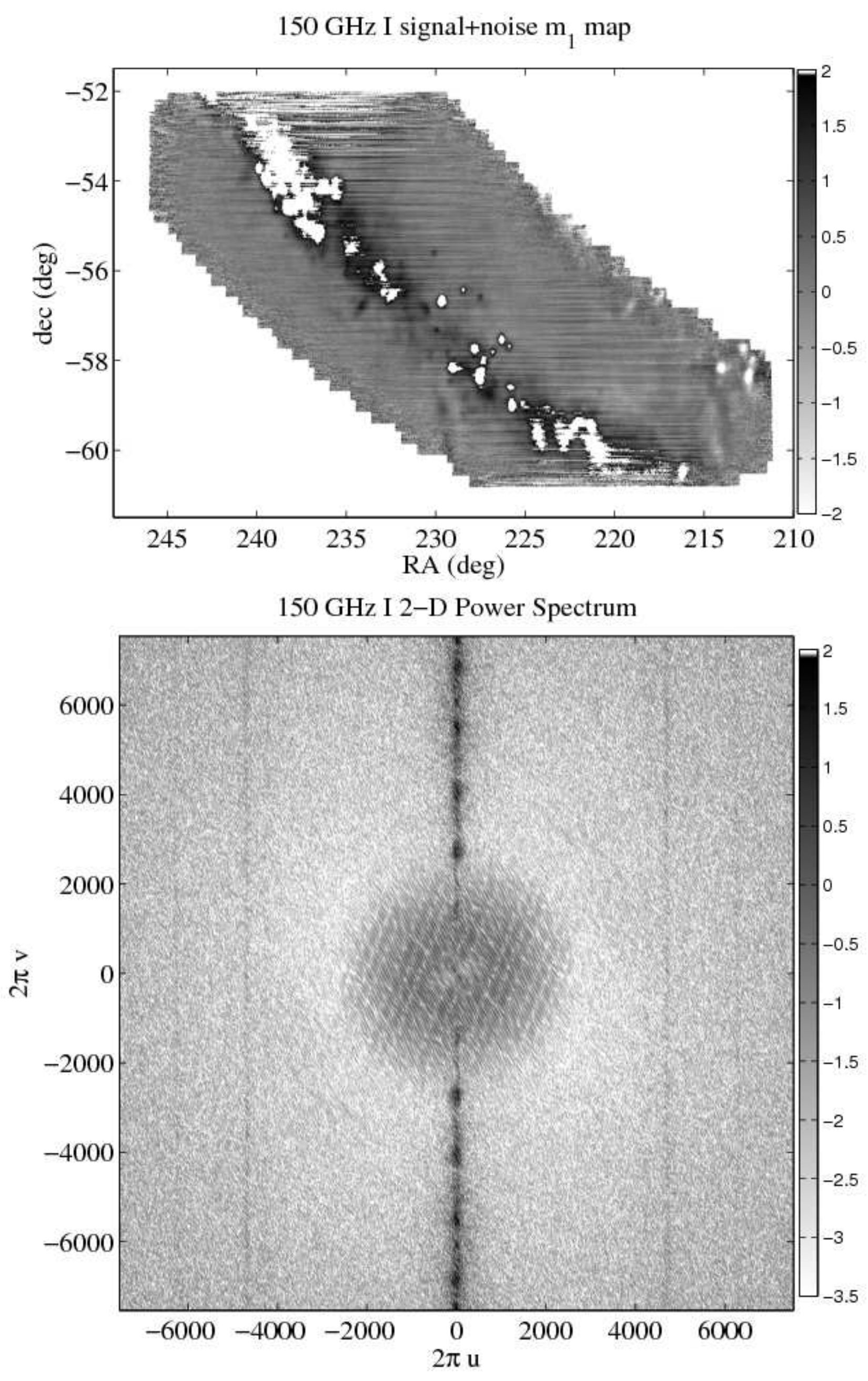}     \\ 
\includegraphics[width=5.5in]{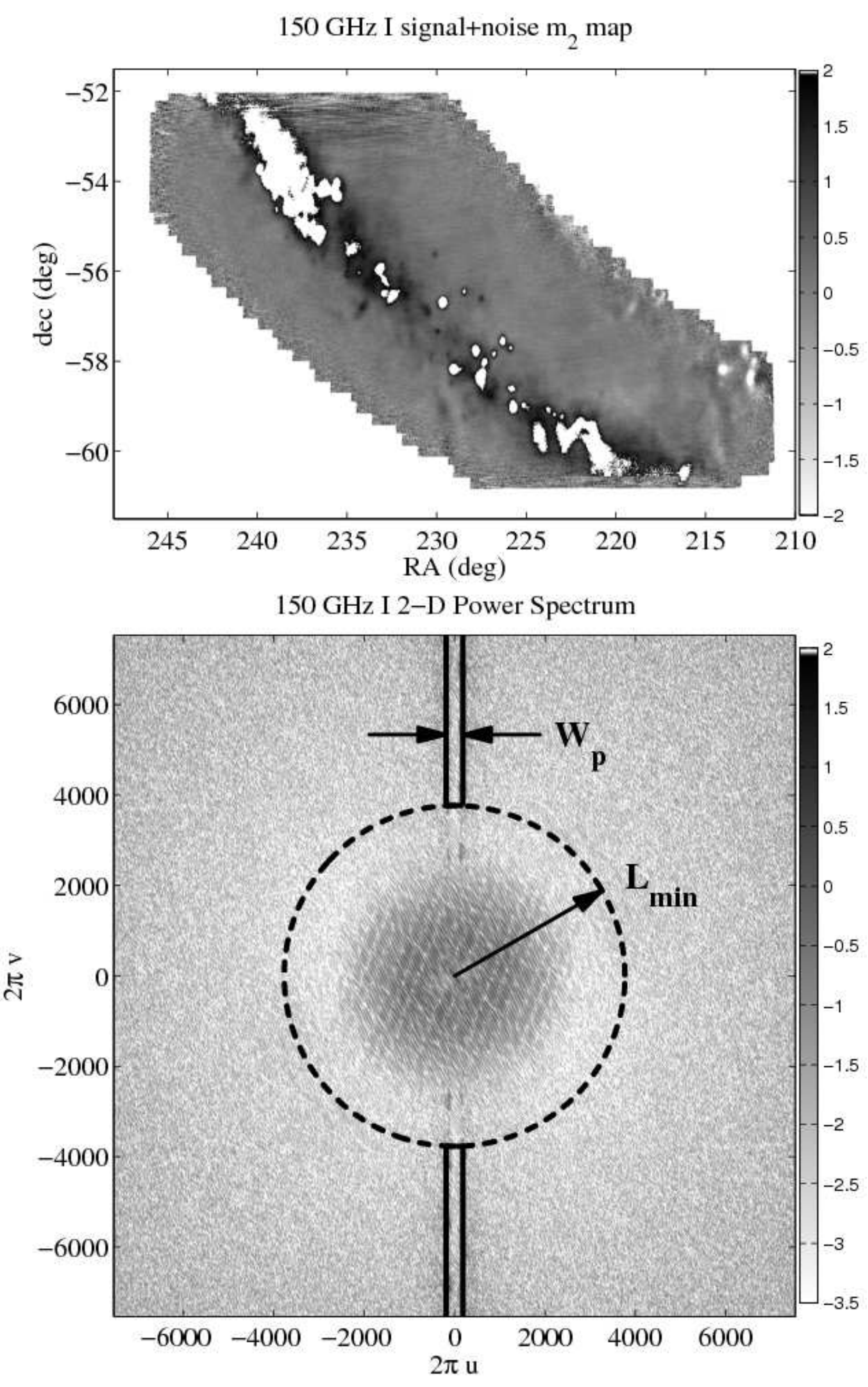}    \\
}
\caption{Demonstration of Fourier plane interpretation of destriping
process using FDS model 8 simulations.
In each column, the \fhigh~\I\ map (top; colorscale in MJy/sr with saturated
 pixels within survey region in white) and corresponding 2D auto 
power spectrum $C_{\bf{u}}$ (bottom) are shown for part of the fourth 
quadrant data.
Note that in the power spectrum plots, the color scale is the
same for each panel, and is logarithmic.
Left: DC-level+slope filtered map $m_{0}$ using
a signal-only simulation. 
In the power spectrum, vertical bands are an unavoidable consequence 
of our scan strategy.
Center: same as Left, for a signal+noise simulation.
The $1/f$ noise, which is added in the scan direction, is obvious
 as the increase in power in the vertical band.
Right: Equivalent plots for the destriped map and $C_{\bf{u}}$ --- 
there is a large reduction in the $1/f$ noise in the destriped map, 
and no obvious residuals have been introduced due to the choice of 
$n_{p}$ and $L_{min}$.
In the power spectrum plot, we show the interpretation of
the trench dug by the polynomial filter of order $n_{p}$ on the
timestream; $L_{min}$ indicates the width of $\tilde{K}$, i.e. 
the angular scale below which modes are preserved. 
Modes inside the circle are subtracted from the timestream, as in
Equation~\ref{eq:noisetod}.
}
\label{fig:mapaps}
\end{figure*}

Experiments such as QUaD which use fixed elevation scans suffer $1/f$ noise
predominantly in the scan direction.
The $1/f$ noise appears as a vertical band in the two-dimensional power 
spectrum $C_{\bf{u}}=\tilde{I}\left(\bf{u}\right)\tilde{I}\left(\bf{u}\right)^{*}$,
where $\tilde{I}$ is the Fourier transform of the total 
intensity map, and ${\bf{u}}=\left(u,v\right)$ is the wavevector.
The leftmost panels of Figure~\ref{fig:mapaps} show a 
signal-only map and $C_{\bf{u}}$ from an FDS simulation of a subsection
 of the survey; an example signal+noise $m_{0}$ map and corresponding 
$C_{\bf{u}}$ are displayed in the center panels, where the vertical noise
band is clearly visible in the 2D power spectrum.
Polynomial filtering over the entirety of each scan digs a `trench' 
into the signal and noise in the vertical band: the width of the trench
 in Fourier space, $W_{p}$, is determined by the polynomial order $n_{p}$, i.e. 
$W_{p}\propto n_{p}$.

In the destriping method, we interpolate `signal-only'
data from the template map $m_{t}$ to remove large-scale (low $|\bf{u}|$)
 modes before filtering, or equivalently removing modes inside
the circle in the center panel of Figure~\ref{fig:mapaps}.
The smallest mode removed is $L_{min}$ --- the width of the Fourier 
transform of the smoothing kernel, $\tilde{K}(\bf{u})$,  used to 
generate the template signal map $m_{t}$. 
With the low-$|\bf{u}|$ modes removed, instead of digging a 
trench along the entire vertical band in the Fourier Plane, as in the
usual case of polynomial filtering, we filter the same width trench 
$W_{p}$, but only for modes with $|\bf{u}|>L_{min}$. 
Residuals from filtering bright point sources (which dominate the signal at
large $\bf{u}$) can be avoided by masking the brightest sources 
during the polynomial filtering in Equation~\ref{eq:sigtod}.

The lower right panel of Figure~\ref{fig:mapaps} demonstrates the 
result when applying the destriping method with a 
6th order polynomial --- filtering has removed much of the noise 
along the vertical axis.
The corresponding map (top right panel) shows
that $1/f$ noise has been heavily suppressed, without the introduction 
of obvious residuals.
We note that some of the $1/f$ noise has been filtered inside 
$L_{min}$; this is due to the fact that $L_{min}$ is not in fact a 
hard boundary in Fourier space, but rather the width of $\tilde{K}$, which 
in the present case is a gaussian function.
Smoothing therefore allows filtering of different modes with 
$\bf{u}$-dependent `weighting', with the weight equal to $\tilde{K}$ evaluated 
at each $\bf{u}$.

The plots indicate that though the algorithm is implemented entirely 
in map/timestream space, its effect is equivalent to a Fourier plane
filter.
Tests on simulated data demonstrate that if uniform scan weighting is 
used, the algorithm is precisely linear in nature.

\section{Effect of Filtering and Map Processing on Diffuse Properties}
\label{app:proprecov}

In the following sections, the recovery of diffuse properties is tested
with signal-only simulations, binned into the same coarse-resolution
pixels used for analysis of the QUaD data.
When comparing input to output values, Equation~\ref{eq:gls} is minimized
as with the real data, except that no error bars are present because signal-only
simulations are used.
All uncertainties quoted in this section therefore reflect the scatter 
introduced by processing effects, and are quantified by the intrinsic
scatter term of Equation~\ref{eq:gls}.

\subsection{Total Intensity Spectral Index}
\label{subsec:specindrecovery}

\begin{figure}[h]
\resizebox{\columnwidth}{!}{\includegraphics{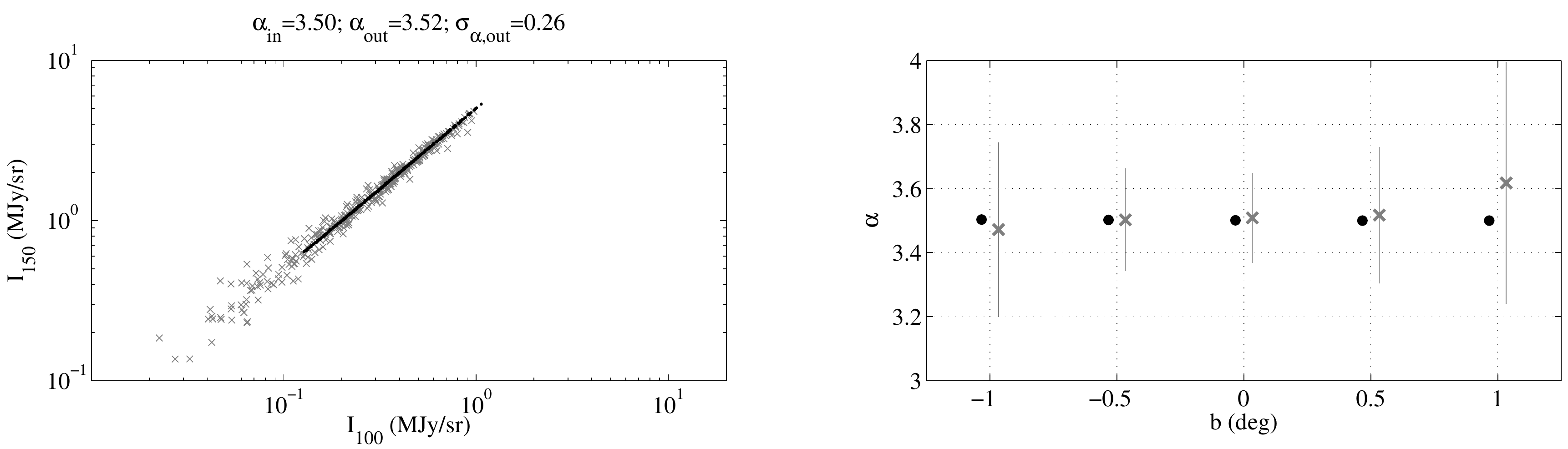}}
\caption{Signal-only simulations of spectral index recovery. Left: Scatter
plot showing input pixel values (black) and those after filtering, field-differencing, 
and destriping (gray crosses). 
The input and output mean spectral index $\alpha$ is shown above the plot,
with the intrinsic scatter introduced by processing effects. 
Right: Spectral index as a function of galactic latitude $b$. 
Mean input is shown as black points, recovered is gray, with error bars indicating
the intrinsic scatter in each $b$ bin.
}
\label{fig:specindrecover}
\end{figure}

To test the recovery of the spectral index, a signal-only simulation of 
the FDS dust model 8 is generated, with the model evaluated at the central 
frequency of each QUaD band as quoted in Section~\ref{subsec:abscal}.
The simulated timestreams are passed through the mapmaking code in
exactly the same way as the real data, thus incorporating the effects
of field-differencing and filtering.

Note that the $m_{2}$ mapmaking stage requires source detection and masking, but 
signal-to-noise thresholds cannot be used to determine the location of 
bright sources with signal only simulations.
To circumvent this problem, signal plus noise simulations of the same simulated 
input maps are generated, with the corresponding $m_{1}$ maps used to determine the
locations of sources to be masked.
The resulting source catalog is then used in the $m_{2}$ mapmaking stage of
the signal-only simulations to reproduce the effects of the filtering
strategy on signal-only data.

The input average spectral index was found to be $\alpha=3.5$, with a 
recovered value of $\alpha=3.52$ and intrinsic scatter $\sigma_{\alpha}=0.26$,
with the scatter introduced by the data processing and mapmaking steps.
The left panel of Figure~\ref{fig:specindrecover} shows the input and recovered
pixel values of $I_{150}$ against $I_{100}$, while the right panel shows 
the input and recovered $\alpha$ as a function of $b$.
Both panels show that the input spectral index is recovered to within the
scatter introduced by the data processing, with a small systematic defecit 
of $<1\%$ in $\alpha$.

\subsection{Polarization}
\label{subsec:polfracrecovery}

The effects of field-differencing and filtering on the recovery of the
polarization fraction are investigated by using two 
signal-only simulations, again with FDS dust model 8 as the input total intensity, 
but using assumed polarization fractions of 2 and 5\%, with the polarization 
signal purely $+Q$ (in galactic coordinates).
From these maps, we simulate QUaD observations and pass the resulting 
timestream through the mapmaking algorithm as in~\ref{app:proprecov}.

\begin{figure}[h]
\resizebox{\columnwidth}{!}{\includegraphics{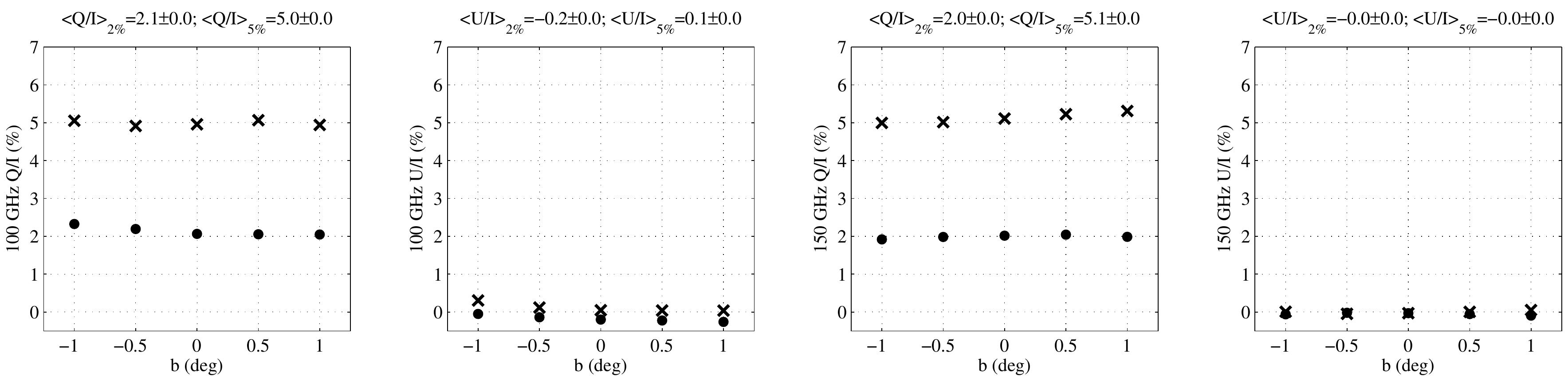}}
\caption{Signal only simulations of galactic-longitude averaged 
polarization fraction, as a function of galactic latitude.
From left to right: \flow~\Qf~and \Uf, \fhigh~ \Qf~and \Uf.
Crosses (dots) correspond to assumed 5 (2)\% polarization fraction 
simulations.
All input polarized signal is in $+Q$ in galactic coordinates, with $U=0$ 
everywhere.
The average recovered values of the polarization fraction are shown above each
plot, with the `error' representing the intrinsic scatter introduced by data 
processing and mapmaking; in the plots of polarization fraction in bins of 
$b$, this quantity is shown as an error bar.
}
\label{fig:polfracrecov}
\end{figure}

Figure~\ref{fig:polfracrecov} demonstrates that for both 2\% and 5\% simulations,
 the input polarization fraction is recovered without introducing any strong 
systematic bias.
Fluctuations in polarization fraction due to filtering and map processing
effects are $0.1$--$0.2$\%, and thus introduce a small error on the 
average polarization fraction measured by QUaD, though such systematic effects
do influence the recovery of polarization angle.

\begin{figure}[ht]
\resizebox{\columnwidth}{!}{\includegraphics{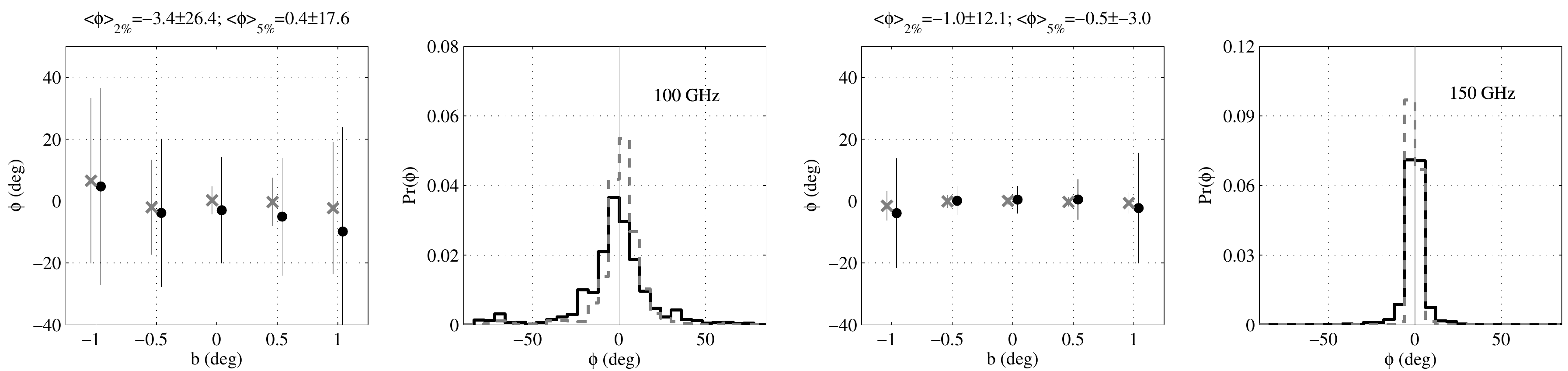}}
\caption{Signal-only simulations of polarization angle $\phi$:
The left two panels are \flow\ $\phi$ vs $b$, and the probability distribution
 of the pixel values as a function of $\phi$.
In the former, the error bars indicate the level of intrinsic scatter introduced
by processing effects.
Gray crosses (black dots) are for the 5 (2)\% polarization fraction simulations.
The numbers above the plots are the average polarization angles in each $b$ 
bin, with the `error' indicating the average intrinsic scatter due to data processing.
In the probability plots, the gray vertical line shows the input $\phi=0^\circ$; gray 
broken (black solid) lines are for the 5 (2)\% polarization fraction simulations.
The right two panels are the same for \fhigh.
}
\label{fig:polanglerecover}
\end{figure}

Figure~\ref{fig:polanglerecover} shows the recovery of polarization angle from the 
same simulations.
Similarly to polarization fraction, the input polarization angle is recovered 
to within 1$\sigma$ of the introduced scatter, both as an average, and within individual bins
 of galactic latitude.
From the average over all pixels, the fluctuation in $\phi$ due to processing 
and filtering the simulated data at \flow~is $17.1^{\circ}$ for 5\% polarization
 fraction, or $26.5^{\circ}$ for 2\% polarization fraction.
At \fhigh, we find  $3.2^{\circ}$ ($13.1^{\circ}$) for 5\% (2\%).
The systematic shift of recovered polarization angle is $\sim5^{\circ}$ ($\sim2^{\circ}$)
for the 2\% polarization fraction simulations at 100 (\fhigh), and 
$\sim0.2^{\circ}$ ($\sim0.4^{\circ}$) for the 5\% polarization fraction simulations
at at 100 (\fhigh).
We conservatively adopt the larger of these quantities when estimating systematic
errors, i.e. the systematic error on the average recovered $\phi$ is $5^{\circ}$
at \flow\ and $2^{\circ}$ at \fhigh.

\section{Celestial Coordinate Maps}
\label{app:maps}

\begin{sidewaysfigure*}
\centering
\resizebox{\textwidth}{!}
{\includegraphics{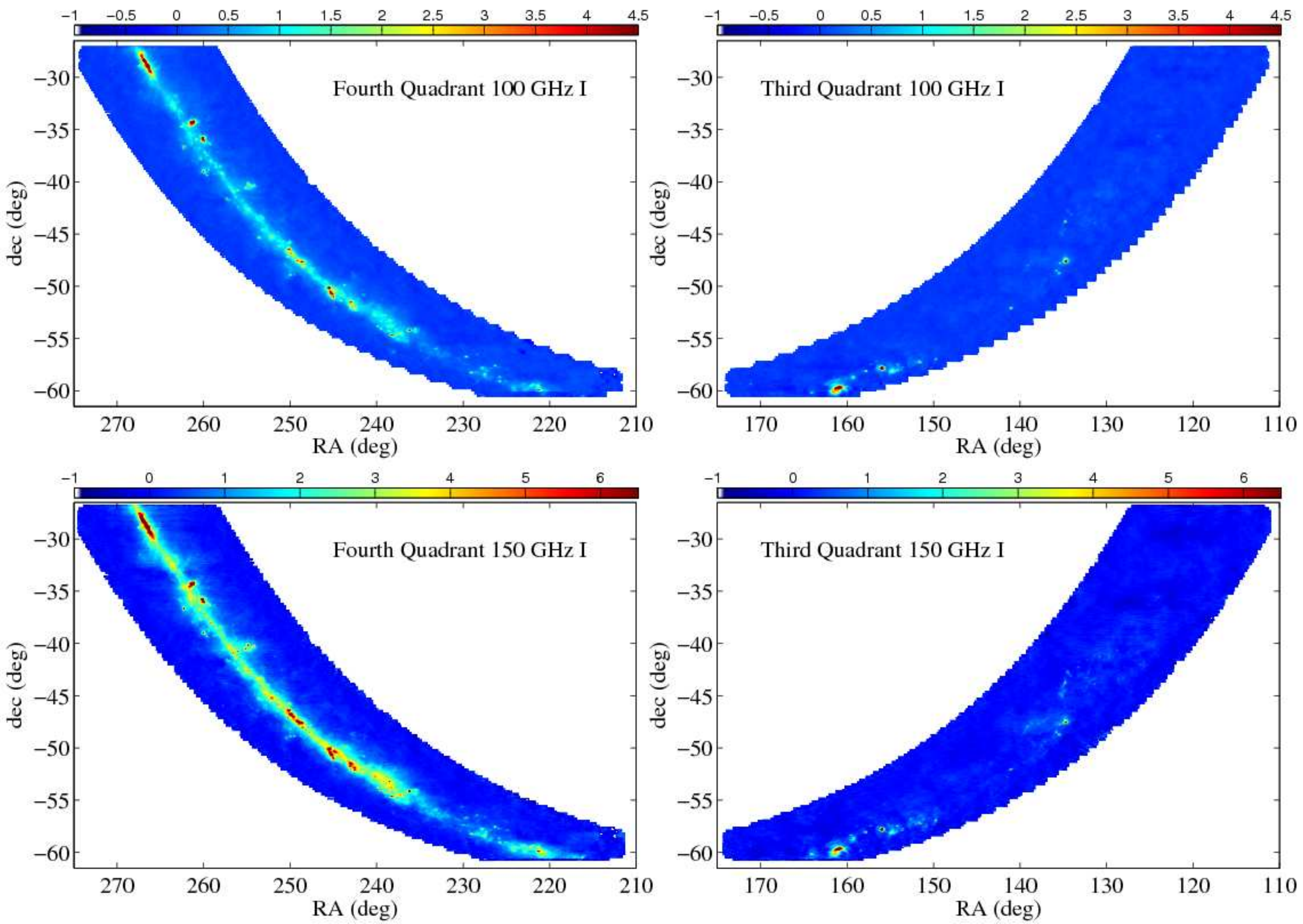}}
\caption{\I~maps for fourth quadrant (left column) and third quadrant 
(right column), at \flow~(top row) and \fhigh~(bottom row). The color scale
is MJy/sr.}
\label{fig:maps_T}
\end{sidewaysfigure*}

\begin{sidewaysfigure*}
\centering
\resizebox{\textwidth}{!}
{\includegraphics{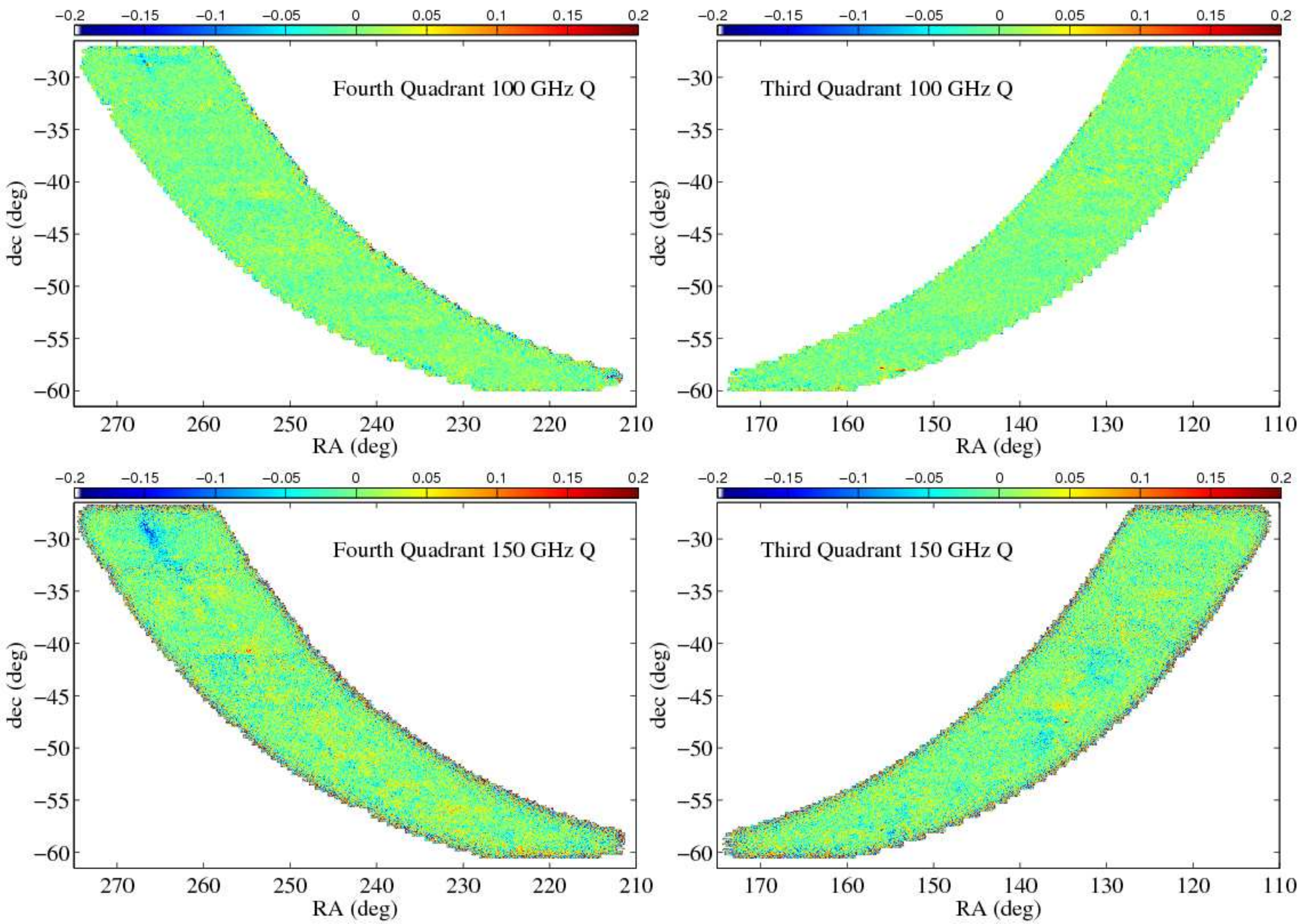}}
\caption{\Q~maps for fourth quadrant (left column) and third quadrant 
(right column), at \flow~(top row) and \fhigh~(bottom row). The color scale
is MJy/sr.}
\label{fig:maps_Q}
\end{sidewaysfigure*}

\begin{sidewaysfigure*}
\centering
\resizebox{\textwidth}{!}
{\includegraphics{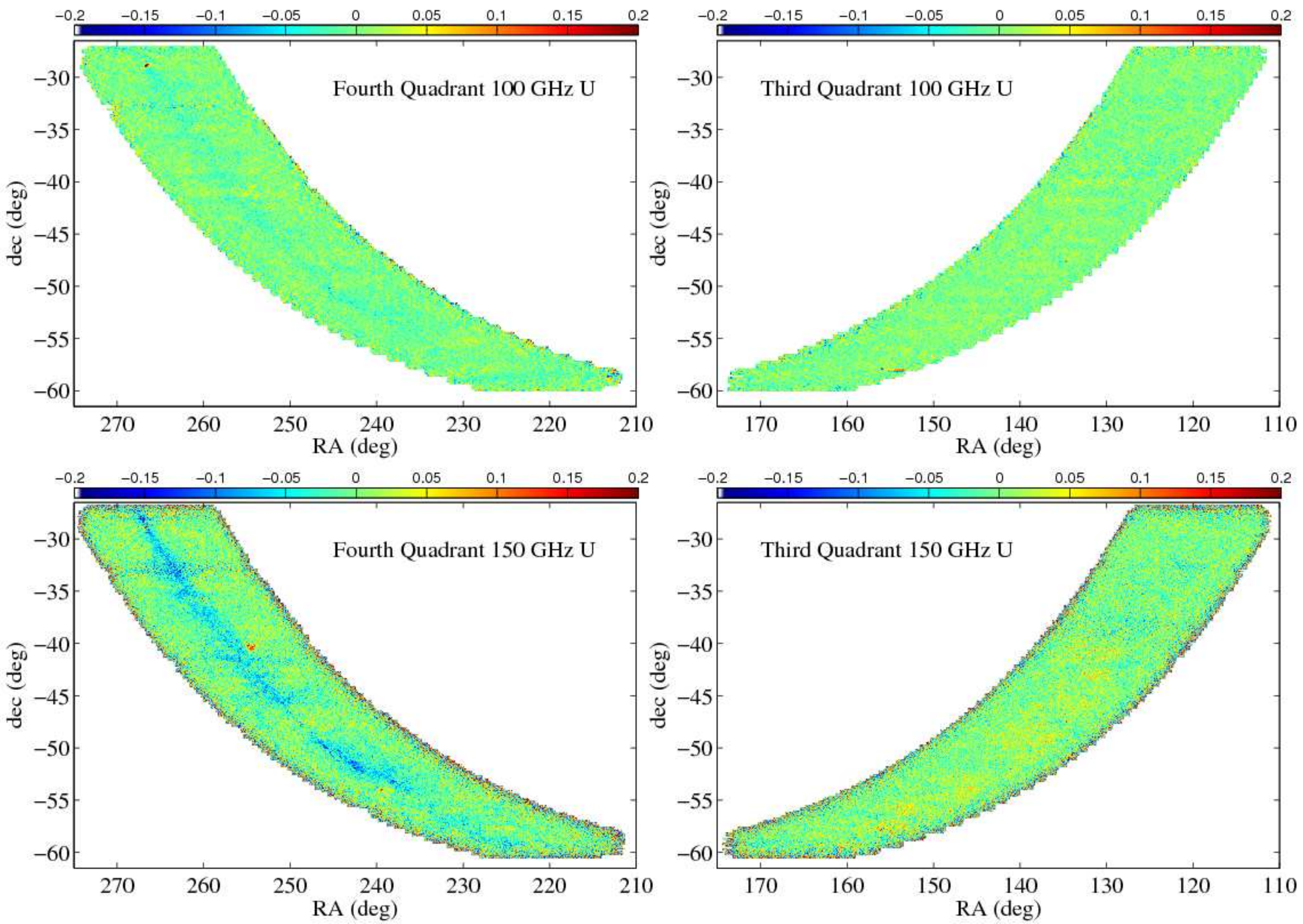}}
\caption{\U~maps for fourth quadrant (left column) and third quadrant 
(right column), at \flow~(top row) and \fhigh~(bottom row). The color scale
is MJy/sr.}
\label{fig:maps_U}
\end{sidewaysfigure*}

Figures~\ref{fig:maps_T}--\ref{fig:maps_U} show the celestial coordinate $m_{2}$ maps.
\clearpage

\end{appendix}

\end{document}